\documentclass[prb,twocolumn,showpacs]{revtex4}
\usepackage{amssymb}
\usepackage{graphicx,amsmath}
\newcommand{\bea}{\begin{eqnarray}}
\newcommand{\eea}{\end{eqnarray}}
\newcommand{\beq}{\begin{equation}}
\newcommand{\eeq}{\end{equation}}
\newcommand{\benu}{\begin{enumerate}}
\newcommand{\enu}{\end{enumerate}}

\newcommand{\al}{\alpha}
\newcommand{\be}{\beta}
\newcommand{\om}{\omega}
\newcommand{\Om}{\Omega}
\newcommand{\ep}{\epsilon}

\newcommand{\si}{\sigma}
\newcommand{\sid}{\sigma^{\dagger}}
\newcommand{\dl}{\delta}
\newcommand{\lam}{\lambda}
\newcommand{\ham}{\mathcal{H}}

\newcommand{\cda}{c^{\dagger}}
\newcommand{\fda}{f^{\dagger}}
\newcommand{\bk}{{\bf k}}
\newcommand{\bq}{{\bf q}}

\begin{document}
\title{Multi-scale fluctuations near a Kondo Breakdown Quantum Critical Point}

\date{\today}
\author{I. Paul$^{1,3}$, C. P\'epin$^2$ and M. R. Norman$^3$}
\affiliation{
$^1$Institut Laue-Langevin, 6 rue Jules Horowitz, B.P. 156, 38042 Grenoble, France \\
$^2$SPhT, CEA-Saclay, L'Orme des Merisiers, 91191
Gif-sur-Yvette, France \\
$^3$Materials Science Division, Argonne National Laboratory, Argonne, IL 60439
}

\begin{abstract}
We study the Kondo-Heisenberg model using a fermionic representation
for  the localized spins. The mean-field phase
diagram exhibits a zero temperature quantum critical point separating a spin
liquid phase where the f-conduction hybridization vanishes, and a Kondo phase
where it does not.
Two solutions can be stabilized in the Kondo phase,
namely a uniform hybridization
when the band masses of the conduction electrons and the f spinons
have the same sign, and a modulated one when they
have opposite sign. For the uniform case,
we show that above a very small  Fermi liquid temperature scale ($\sim$ 1 mK),
the critical fluctuations associated with the vanishing hybridization
have dynamical exponent $z=3$, giving rise to a specific heat coefficient
that diverges logarithmically in temperature, as well as a conduction
electron inverse lifetime that has a $T \log T$ behavior.  Because the
f spinons do not carry current,
but act as an effective bath for the relaxation of the current carried by the conduction electrons,
the latter result also gives rise to a
 $T \log T$ behavior in the resistivity.  This behavior is consistent
 with observations in a number of heavy fermion metals.
\end{abstract}

\maketitle

\section{Introduction}
\label{sec:intro}

A large number of experiments have been performed on metallic
 heavy fermion compounds close to a  zero temperature phase transition
 (a quantum critical point (QCP))
driven by applied magnetic field, chemical doping or
pressure \cite{stewart,hilbert}. In the quantum critical regime,  the
thermodynamics and transport properties indicate a breakdown
of the Fermi liquid. In many cases, the resistivity is quasi-linear
in temperature over several decades, and
the specific heat coefficient diverges logarithmically.
The spin
susceptibility typically exhibits
 an anomalous exponent in temperature.
 Neutron scattering experiments on some
 of these materials have
  revealed that the anomalous exponent in the
dynamical susceptibility is identical for all points in the
Brillouin zone~\cite{ray,almut}, suggesting a local character for the fluctuations.
deHaas-vanAlphen experiments also find a divergence of the effective mass
when approaching the critical point, along with a change in the Fermi surface
topology when going through it~\cite{dhva}.

These unusual observations have motivated many
theoretical studies that have attempted to capture these effects.
Most theories~\cite{hertz,millis,moriya,rosch,hilbert}
are based on the assumption that at the
QCP,  a spin density wave forms,
and therefore the critical fluctuations that destabilize the Fermi liquid are
magnetic in nature~\cite{hertz,millis,moriya,rosch,hilbert}.
In three dimensions, these
theories fail to capture simultaneously the linear
temperature dependence of the
resistivity, the logarithmic divergence of the specific heat coefficient~\cite{review-piers},
 and the anomalous exponent of the spin
susceptibility~\cite{pankov}.
For an antiferromagnetic spin density wave transition, a central problem is that the
critical fluctuations are confined to an inverse coherence length about the spin density
ordering vector, and consequently, only parts of the Fermi surface couple effectively with the
critical bosonic modes.

More recently, the problem has been approached from another
perspective which takes the point of view that
at the QCP,
magnetic fluctuations  suppress the formation of the heavy Fermi liquid,
driving the effective Kondo temperature of the
lattice ($T_K$) to zero~\cite{review-piers,qimiao,senthil,schofield}.
In this picture, the QCP is a fractionalized critical point at which the heavy
quasiparticle deconfines into a spinon and holon.
One feature that distinguishes between these two classes of theories is that
the first predicts the Fermi surface to change smoothly across
the QCP, while the second predicts an abrupt change~\cite{review-piers}.
Recent results of the Hall effect
for YbRh$_2$Si$_2$~\cite{paschen}, as well as the earlier mentioned dHvA data~\cite{dhva},
have lent support to theories of the second type.

Here, we explore the possibility that
in the quantum critical regime,
the magnetic
fluctuations are not the dominant ones at the QCP, and that
the unusual
behavior in thermodynamics and transport is due to
critical fluctuations of a non-magnetic order parameter
associated with the vanishing energy scale $T_K$.
One motivation for this point of view is the fact  that
in some compounds like YbRh$_2$Si$_2$, the gain in entropy inside
the magnetically ordered phase represents only a few percent of
the total entropy per localized spin~\cite{custers}.
 The order parameter we
advocate is the field $\sigma$ associated with the hybridization between
the localized
spins and the conduction electrons~\cite{read,millis-lee}.
At the QCP,  the effective
Kondo temperature for the lattice goes to zero, leading
to a `Kondo breakdown' of the heavy Fermi liquid.
The critical fluctuations of $\sigma$ are gapless excitations,
and we study how these fluctuations influence
the properties of the metal using
 the formalism of the large $N$ Kondo-Heisenberg model.

There have been several earlier studies of this model~\cite{antoine,senthil,schofield}.
 Beyond the mean-field level, the
Kondo-Heisenberg model can be treated as a lattice gauge theory.
 Senthil {\it et.~al.}~\cite{senthil} have examined the effect of the gauge fluctuations in
 this model, while
Coleman {\it et.~al.}~\cite{schofield} studied the zero temperature transport
 anomalies.  In our work, we find a number of novel effects associated with
 the fluctuations of the $\sigma$ field which were not discovered in these earlier studies.

At the Kondo breakdown  QCP, the metal passes from a magnetic phase (which we approximate,
as in earlier work~\cite{senthil}, as a uniform spin liquid) to a Kondo phase.
In the spin liquid phase, the f spinons are characterized by a `Fermi surface' which generically differs
in size from the conduction electron Fermi surface.  In the Kondo phase, these two surfaces
become coupled due to the non-zero expectation value of $\sigma$.
In our study, we observe two new phenomena associated with this.
First, for the case where the spinon and conduction electron masses have opposite sign,
$\sigma$ can order at a finite wavevector,
leading to spatial modulations of the Kondo hybridization
analogous to the LOFF state  of superconductivity~\cite{fflo,rice}.
Second, we find the presence of multiple energy scales, spread over a very large
range in energy, due to the mismatch between the two Fermi surfaces.
The lowest scale, below which Fermi liquid behavior is restored, is extremely small (of order 1 mK),
above which, up to an ultraviolet cutoff of order the single ion Kondo temperature,
the critical fluctuations of $\sigma$ exhibit a
dynamical exponent $z=3$.
This gives rise to a marginal Fermi liquid like
behavior in $d=3$ for the conduction electrons along the entire
Fermi surface, due to scattering with the critical fluctuations.
This property is to be contrasted with antiferromagnetic spin density wave models,
where only on parts of the Fermi surface the scattering of the electrons with the
critical mode is effective.
Next, since the f spinons do not carry current,
but act as an effective bath for the relaxation of the current carried by the conduction electrons,
the marginal Fermi liquid behavior also gives rise to
a resistivity that goes as $T \log T$.  This behavior is unlike  either that of
ferromagnetic spin density wave
models in which the transport lifetime is less singular than the single particle lifetime
(i.e., in the latter models, forward scattering does not degrade the current),
or that of antiferromagnetic spin density wave models in which the ``cold'' parts of the
Fermi surface dominate the transport properties~\cite{hlubina}.
Moreover, a logarithmic dependence is found
for the specific heat coefficient from both the gauge~\cite{senthil}
and $\sigma$ fluctuations.  The latter also give rise to an anomalous temperature
 exponent of 4/3 in the uniform spin susceptibility.
 A summary of our results have been presented in a shorter paper~\cite{prl}.

The phenomenon of the breakdown of the Kondo effect at a QCP can also be studied
in the more general context of a periodic Anderson model.
This generalization is discussed in other works~\cite{anderson,lorenzo}.

\section{Model and Formalism}

The starting point of our theory is the microscopic Kondo-Heisenberg model
in three dimensions,
which describes a broad band of conduction electrons interacting with a
periodic array of localized spins through antiferromagnetic Kondo coupling
$J_K > 0$. Additionally, the localized spins interact with one another via
nearest neighbour exchange $J_H > 0$. The Hamiltonian for the large $N$
version of this model, where $N$ denotes the enlarged spin symmetry group
$SU(N)$, is given by
\bea
\label{eq:hamiltonian}
\ham &=& -t \sum_{\langle ij \rangle, \al}  \cda_{i \al} c_{j \al}
+ \frac{J_K}{N} \sum_{i, \al, \be} \cda_{i \al} c_{i \be} \fda_{i \be} f_{i \al}
\nonumber \\
&+&
\frac{J_H}{N} \sum_{\langle ij \rangle, \al, \be}
\fda_{i \al} f_{i \be} \fda_{j \be} f_{j \al}.
\eea
Here $\cda_{i \al}$ ($c_{i \al}$) are creation (annihilation) operators for the
conduction electrons
with spin index $\al = (1,N)$ at site $i$, and $\langle ij \rangle$ refers to
nearest neighbour sites. $t$ is the hopping matrix element between neighbouring
sites for the conduction electrons.
The $SU(N)$ generalization of the localized spins $S^{a}_i$ with
$a = (1, ..., N^2-1)$ at each site
$i$ are expressed in terms of Abrikosov pseudofermions (or spinons) by
$S_i^{a} = \sum_{\al \be} \fda_{i \al} (\Gamma^{a}_{\al \be}/N) f_{i \be}$,
where $\Gamma^{a}$ are the generators of the $SU(N)$
group in the fundamental representation. This fermionic
representation of the spin operator gives rise  to a local constraint at each site, $i$
\beq
\label{eq:constraint}
\sum_{\al} \fda_{i \al} f_{i \al} = \frac{N}{2},  \qquad \forall i.
\eeq
We note that in the context of the heavy fermion systems, the
Heisenberg exchange term is often equated to the RKKY interaction between
the localized spins which is mediated by the mobile conduction electrons.
In such a scenario, the Heisenberg coupling $J_H \propto \rho_0 J_K^2$,
where $\rho_0$ is the density of states of the conduction electrons at the
Fermi level. However, for the purpose of the present study, it is convenient to
consider $J_H$ as a parameter independent of $J_K$. Microscopically this can be
justified by noting that, in principle, there can be other sources which generate
this coupling, such as superexchange within the narrow band of
$f$-electrons.

In order to perform a systematic large $N$ study of the system defined by
Eqs.~(\ref{eq:hamiltonian}) and (\ref{eq:constraint}), the first step is to decouple
the interaction terms which are quartic in fermionic operators using a Hubbard-Stratonovich
transformation. The Heisenberg exchange term is decoupled using a
bosonic link variable $\phi_{ij} \rightarrow \sum_{\al} \fda_{i \al} f_{j \al}$, while
the Kondo interaction is decoupled by introducing a complex bosonic field
$\sid_i \rightarrow \sum_{\al} \fda_{i \al} c_{i \al}$.
In the next step, following Ref.~\onlinecite{senthil}, we assume that in three dimensions,
$\phi_{ij}$ condenses in a uniform spin liquid phase, i.e.,
$\langle \phi_{ij} \rangle = \phi_0$ at the mean field level. This provides a dispersion
to the spinon band which, as we show later, is an essential ingredient to obtain the
breakdown of the Kondo effect.
We note that there is no clear evidence of a spin liquid phase in any
heavy fermion system near its quantum critical point. Rather, the typical phase diagram exhibits a QCP that
separates a magnetic ground state (typically an antiferromagnet) from a paramagnetic
heavy Fermi liquid. Consequently,
it is useful to discuss the motivation
for our choice of a uniform spin liquid phase for the Heisenberg link variable
$\phi_{ij}$.
This choice is partly guided by convenience: since our main purpose is to
study the consequences
of the breakdown of the Kondo effect, the choice of a uniform spin liquid can be
viewed as the
simplest device which allows the vanishing of the Kondo energy scale (indicating the
breakdown
of the Kondo effect) at the mean field level. More physically, one can view the uniform
spin liquid
as a mean field description of the short range magnetic correlations that persist
when a magnetic
ground state is destroyed by quantum fluctuations. However, to demonstrate
this point concretely
is not simple, and beyond the scope of the present study.
The key point is that the spin liquid provides a bandwidth for the f electrons.
Other approaches, for instance one where the bandwidth is due to direct f-f hopping,
should yield similar results in regards to the breakdown of the Kondo effect that we
describe here.

The system can now be described by the Lagrangian
\bea
\label{eq:lagrangian}
\mathcal{L}   &=&
\sum_{\langle ij \rangle \alpha}  \left[ c^{\dagger}_{i
\alpha}  \left(  \partial_{\tau} \dl_{ij} -t \right) c_{j \alpha} +
f^\dagger_{i   \alpha}  \left ( \left(
\partial_{\tau} - \lambda_i \right) \dl_{ij}
\right. \right.
\nonumber \\
&-&
\left. \left.
\phi_0 e^{i a_{ij}} \right) f_{j \alpha} \right]
+ \frac{N}{2} \sum_i \lambda_i + \frac{N}{J_K} \sum_i \sigma^{\dagger}_i \sigma_i
+ \frac{N\phi_0^2}{J_H} \nonumber \\
&+&
 \sum_{i \alpha} \left( c^\dagger_{i \alpha} f_{i \alpha} \sigma_i
+ \rm{H.c.} \right),
\eea
where $V$ (the volume of the system) is set to 1.
In the above,  $\lam_i$ are Lagrange multipliers (scalar potential) that
enforce the local constraint of $N/2$ spinons per site. Now, given a
many-body wave function that satisfies this constraint, a single hop of a
spinon takes the state out of the physical subspace. Consequently, for the
kinematics of the spinons, only simultaneous opposite hops between two
neighbouring sites is a physically allowed process. This implies that the
local spinon current operator $\vec{J}_{f i} = 0$ at every site $i$.
The gauge fields $a_{ij}$ (vector potential), associated with the phase of
$\phi_{ij}$, ensure that this constraint is satisfied.
The appearance of the scalar and vector potentials can also be understood
by noting that $\mathcal{L}$ is invariant (up to a term which is a total
derivative of imaginary time) under a local $U(1)$ gauge
transformation $f_{i \al} \rightarrow f_{i \al} e^{i \theta_i}$,
$\sigma_i \rightarrow \sigma_i e^{-i \theta_i}$,
$\lam_i \rightarrow \lam_i + i \partial_{\tau} \theta_i$,
$a_{ij} \rightarrow a_{ij} - \theta_i + \theta_j$,
a consequence of the fermionic representation of the spin and the
constraint Eq.~(\ref{eq:constraint})~\cite{ioffe}.

\begin{figure}[tbp]
\begin{center}
\includegraphics[width=2.4in]{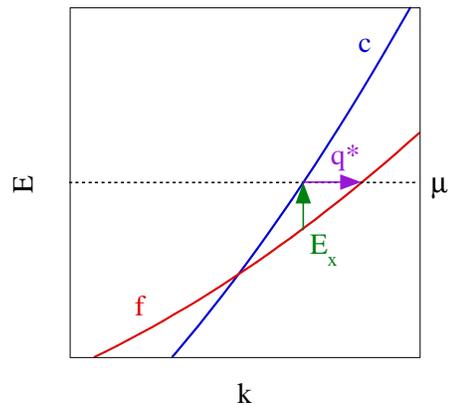}
\end{center}
\caption{(Color online) Dispersion of conduction and spinon bands, with the mismatch wavevector, $q^*$,
and the mismatch energy, $E_x \equiv \alpha v_F q^*$, indicated, where $\alpha$ is the ratio of the
spinon and conduction bandwidths.  An artificially large value of
$\alpha$ was used in this plot (0.5) so as to better illustrate the origin of $E_x$.}
\label{fig:band}
\end{figure}

In the following we examine the above Lagrangian, first in a mean
field approximation
and then
consider Gaussian fluctuations of the action around the mean field
solution. This involves studying the possibility
of hybridization between the conduction and the spinon bands (for
$\langle \si_i \rangle \neq 0$) as well as calculating the hybridization
fluctuation which is an interband particle-hole excitation. As
such, one needs to characterize the dispersions of the conduction and
the spinon bands. We do this by assuming that the bands have a parabolic
dispersion (to facilitate calculations), and we introduce the following
 two important parameters.
 First, $\alpha \equiv \phi_0/D$, is the ratio of
the spinon bandwidth $\phi_0$ and the conduction bandwidth $D$. As
we will see in the next section, at the Kondo breakdown QCP $\phi_0
\sim J_H \sim T_K^0$, where $T_K^0 \equiv D e^{- 1/(\rho_0 J_K)}$ is
the single-ion Kondo energy scale of the system, which is typically of order
10 K in heavy fermion systems. Assuming $D \sim 10^4$ K, we get
$\al \sim 10^{-3}$. Second, while the spinon band is half filled due
to the constraint (for $N=2$), the conduction band
filling is generic. Without any loss of generality, we take the
conduction band to be less than half filled. This implies that the
Fermi wave vector of the conduction band $k_F$ is different from
that of the spinon band $k_{F0}$. We denote this mismatch by
$q^{\ast} \equiv k_{F0} - k_F$, and assume that the fraction
$(q^{\ast}/k_F)$ is of the order 0.1~\cite{foot1}. This would mean that while
$k_F$ and $k_{F0}$ are of the order of the Brillouin zone dimension,
the mismatch wave vector $q^{\ast}$ is one order of magnitude
smaller. The parameters $\alpha$ and $(q^{\ast}/k_F)$ affect the
important energy scales of the system.  This is illustrated in Fig.~\ref{fig:band}
where we show the conduction and spinon dispersions.

\section{Mean Field Treatment}
\label{sec:meanfield}

At the level of the mean field approximation, we replace the bosonic
Hubbard-Stratonovich fields and the Lagrange multipliers by their
expectation values, and we study the approximate Lagrangian given by
\bea
\label{eq:mf-lagrangian}
\mathcal{L}_{MF}   &=&
\sum_{\langle ij \rangle \alpha}  \left[ c^{\dagger}_{i
\alpha}  \left(  \partial_{\tau} \dl_{ij} -t \right) c_{j \alpha} +
f^\dagger_{i   \alpha}  \left ( \left(
\partial_{\tau} - \langle \lambda_i \rangle \right) \dl_{ij}
\right. \right.
\nonumber \\
&-&
\left. \left.
\phi_0  \right) f_{j \alpha} \right]
+ \frac{N}{2} \sum_i \langle \lambda_i \rangle
+ \frac{N}{J_K} \sum_i \left| \langle \sigma_i \rangle \right|^2
+ \frac{N\phi_0^2}{J_H} \nonumber \\
&+&
 \sum_{i \alpha} \left( c^\dagger_{i \alpha} f_{i \alpha} \langle \sigma_i \rangle
+ \rm{H.c.} \right).
\eea
In the following we write the dispersion ($\ep_{\bk}$) of the conduction band as
\beq
\label{eq:cond-dispersion}
\ep_{\bk} = \ep + \frac{\ep^2}{D},
\eeq
where $\ep = v_F (k - k_F)$, $k$ is the magnitude of ${\bf k}$,
and $v_F$ is the Fermi velocity of the conduction
electrons. The dispersion ($\ep^0_{\bk}$) of the spinon band is similarly written as
\beq
\label{eq:f-dispersion}
\ep^0_{\bk} = \al \left[ (\ep - v_F q^{\ast}) + \frac{(\ep - v_F q^{\ast})^2}{D} \right].
\eeq
We note that, in the above, both the bands are taken as electron-like, for which we find
that the mean field equations yield a spatially uniform solution, namely,
$\langle \si_i \rangle = \si_0$ and $\langle \lam_i \rangle = \lam_0$. In the case where
one of the bands is chosen to be hole-like, we find a spatially modulated solution which
we discuss in appendix \ref{appensub:modulated}.
The free energy corresponding to Eq.~(\ref{eq:mf-lagrangian})
is given by
\bea
\label{eq:mf-free1}
\frac{F_{MF}}{N}
\! &=& \!
- \frac{1}{\be}  {\rm Tr} \left[ \ln \left(- G^{-1}_a (i \om_n, \bk) \right)
+ \ln \left( - G^{-1}_b (i \om_n, \bk) \right) \right]
\nonumber \\
&+&
\frac{\si_0^2}{J_K} + \frac{\phi_0^2}{J_H}  + \frac{\lam_0}{2},
\eea
where $\be$ is the inverse temperature, $\om_n$ is the fermionic Matsubara frequency,
and Tr corresponds to a trace over space-time co-ordinates. In the above
\beq
\label{eq:G-ab}
G^{-1}_{a,b}(i \om_n, \bk)  = i \om_n - \ep_{\bk}^{ a,b},
\eeq
where
\beq
\label{eq:epsilon-ab}
\ep_{\bk}^{a,b} = \frac{1}{2} \left[ \ep_{\bk} + \ep^0_{\bk}
\mp \sqrt{(\ep_{\bk} - \ep^0_{\bk})^2 + 4 \si_0^2} \right].
\eeq
We evaluate the free energy
given by Eq.~(\ref{eq:mf-free1}) at zero
temperature ($T=0$) in the limit $(q^{\ast}/k_F) \rightarrow 0$. The details of this evaluation is given
in appendix \ref{appensub:free}. As a function of $\al$ and $\si_0$, and to $\mathcal{O}(\si_0^4)$ accuracy,
we find ($\alpha \ll 1$)
\bea
\label{eq:mf-free2}
\frac{F_{MF}}{N}
&=&
\frac{\rho_0 D^2}{2} \left[
\frac{\al^2}{2 \rho_0 J_H} - \frac{\al}{3} \right]
+ \rho_0 \si_0^2 \left[ \frac{1}{\rho_0 J_K}
\right. \nonumber \\
&-&
\left.
\ln \left( \frac{1}{\al} \right) \right]
+ \frac{\rho_0 \si_0^4}{\al^2 D^2} + {\rm const},
\eea
where the constant part has explicit $\lam_0$ dependence.
Since the precise value of $\lam_0$ is of no importance for our results, in the following
we ignore the mean field equation for $\lam_0$. Minimizing $F_{MF}$ with
respect to $\al$ and $\si_0$ we get
\beq
\label{eq:mf-alpha}
\frac{\rho_0 D^2}{2} \left[ \left( \frac{\al}{\rho_0 J_H} - \frac{1}{3} \right)
+ \frac{2 \si_0^2}{\al D^2} - \frac{4 \si_0^4}{\al^3 D^4} \right] = 0,
\eeq
\beq
\label{eq:mf-sigma0}
2 \rho_0 \si_0 \left[ \left( \frac{1}{\rho_0 J_K} - \ln \left( \frac{1}{\al} \right) \right)
+ \frac{2 \si_0^2}{\al^2 D^2} \right] = 0,
\eeq
respectively. We study these equations by keeping the Heisenberg parameter $J_H$ fixed, while
varying the Kondo parameter $J_K$, and find two solutions corresponding to two mean
field ground states. (i) First, a uniform spin liquid phase where $\si_0 = 0$, which implies
that in this phase, the Kondo effect fails to occur and the localized spins
remain unscreened in a uniform spin liquid state. In this phase, $\al = \al_0 \equiv (\rho_0 J_H)/3$, which
implies that the Heisenberg coupling sets the scale for the spinon dispersion, since
$\phi_0 = (\rho_0 D J_H)/6 \sim J_H$. It is simple to check that this solution is stable for
$J_K < J_{K_c}$, where
\beq
\label{eq:JKc}
\frac{1}{\rho_0 J_{K_c}} = \ln \left( \frac{1}{\al} \right).
\eeq
(ii) For $J_K > J_{K_c}$ the stable mean field solution corresponds to $\si_0 \neq 0$, indicating
a ground state where the local moments are screened by the Kondo effect and a heavy Fermi
liquid is established below an energy scale $T_K \approx \pi \rho_0 \si_0^2$. The growth of the Kondo
order parameter in this phase is given by
\beq
\si_0 \propto J_H \ln \left( \frac{1}{\al_0} \right) \left[ \frac{J_K - J_{K_c}}{D} \right]^{\be},
\eeq
where $\be = 1/2$ is the typical mean field exponent. We also find that the spin liquid
order parameter decreases in this phase, and is given by
\beq
\label{eq:mf-alpha2}
\al = \al_0 - \frac{6 \si_0^2}{D^2} + \mathcal{O} (\si_0^4).
\eeq

Thus, from the above mean field study, we find that, in the presence of a finite bandwidth of the
spinons, the Kondo effect takes place only when the Kondo coupling $J_K$ is larger than a finite
value $J_{K_c}$. This establishes the Kondo breakdown QCP where the lattice Kondo energy scale
$T_K$ vanishes. In the current formulation of the mean field theory, the Kondo breakdown QCP
separates a uniform spin liquid ground state ($J_K < J_{K_c}$) from a heavy Fermi liquid ground state
($J_K > J_{K_c}$). It is important to note that if we define a single-ion Kondo scale ($T_K^0$)
as a function of $J_K$ for the system
by
\beq
T_K^0 (J_K) \equiv D e^{-1/(\rho_0 J_K)},
\eeq
using Eq.~(\ref{eq:JKc}) we conclude that at the QCP
\beq
\label{eq:JhTk}
J_H \sim T_K^0 (J_{K_c}).
\eeq
This shows that the Kondo breakdown QCP is established as a result of a competition between
the Kondo energy scale and the magnetic energy scale, even though there is no long range magnetic
order in the present study.
The reduction of the spin liquid order parameter, given by Eq.~(\ref{eq:mf-alpha2}), provides
further evidence for this competition.
Therefore, this mean field study can be viewed as a microscopic
realization of the energetic argument that Doniach had proposed several decades ago for the
existence of a QCP in heavy fermion systems~\cite{doniach}.

\section{Fluctuations}
\label{sec:fluctuations}

In this section, we study the massless fluctuations in the
quantum critical regime. There are two such modes: (a) one associated with the
phase of $\phi_{ij}$ which are the gauge fluctuations, and (b) the
fluctuations of the complex order parameter $(\si^{\dagger}_i, \si_i)$
which are gapless due to the vanishing of the Kondo energy
scale $T_K$ at the Kondo breakdown QCP.

\subsection{Gauge Fluctuations}
\label{subsec:gauge}

Since the gauge fluctuations of the system have been studied earlier~\cite{senthil}, here we just
summarize the main points for the sake of completeness. It is convenient to work
in the Coulomb gauge $\vec{\nabla} \cdot \vec{a} = 0$, where the vector gauge fields
$a_{\mu}$ $(\mu = x, y, z)$ are purely transverse~\cite{ioffe}. In this gauge the fluctuations of
the scalar potential $\lambda$ decouple from $a_{\mu}$, and give rise to a screened
Coulomb interaction between the spinons which can be neglected. Next, since the fields
$a_{\mu}$ enter the theory as vectorial Lagrange multipliers to satisfy the
constraint that the local spinon current is zero, they behave as `artificial photons'
without any intrinsic dynamics of their own. Their dynamics is entirely generated by
their coupling to the matter field, namely the spinon band, and therefore these bosonic
modes are overdamped. The propagator for the transverse gauge fields is defined
as $D_{\mu \nu} ({ \bf x}, \tau ) = \langle T_{\tau} \left
[a_\mu ({\bf x}, \tau ) a_\nu (0,0) \right ] \rangle$, which in frequency-momentum
space has the standard form
$D_{\mu \nu} ( \bq, i \Omega_n) = ( \delta_{\mu \nu} -
q_\mu q_\nu/ q^2 ) \Pi^{-1} (q, i \Omega_n)$, with $\Pi (q, i
\Omega_n) \propto [ (q/2k_{F0})^2 + |\Omega_n|/(\alpha v_F q)]$.
Here $\Om_n$ is a bosonic Matsubara frequency, and the above expression for the
gauge propagator $D_{\mu \nu}(\bq, i \Om_n)$ is valid for frequencies smaller
than the spinon bandwidth $\alpha D$. As a result, the gauge excitations are
characterized by a dynamical exponent
$z=3$, which  in $d=3$ are known~\cite{reizer} to give a contribution to the
specific heat coefficient $\gamma \equiv - \partial^2F/\partial
T^2 \propto \ln(\alpha D/T)$ and to the static spin susceptibility
$\dl \chi_s \propto T^2 \ln (\al D/T)$.
Finally, it has been argued in the
literature that
the gauge fluctuations convert the finite temperature mean field phase transition
line into a crossover line~\cite{senthil,nagaosa-lee}.

\subsection{Fluctuations of the Kondo Boson}
\label{subsec:kondofluc}

\begin{figure}[tbp]
\begin{center}
\includegraphics[width=8.8cm]{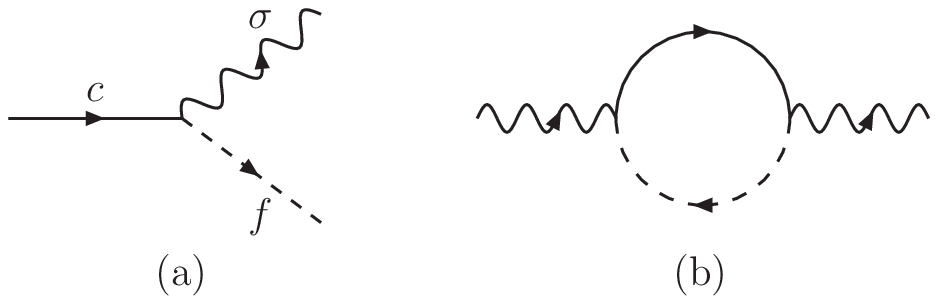}
\end{center}
\caption{(a) Vertex for the interaction between the conduction electrons (solid line) and the
spinons (dashed line) mediated by the hybridization fluctuations $\sigma$ (wiggly line).
(b) The interband polarization involving conduction electrons and spinons,
which generates the dynamics of $\sigma$. For momentum transfer
$q > q^{\ast}$, where $q^{\ast}$ is the mismatch between the conduction and the spinon Fermi
surfaces, $\sigma$ is an overdamped critical mode with dynamical exponent $z=3$. }
\label{fig:polarization}
\end{figure}

At the QCP, where the Kondo coupling is tuned to its critical value $J_{K_c}$,
the critical fluctuations of the continuous phase transition are given by those
of the complex order parameter fields $(\si^{\dagger}, \si)$.
The propagator for these fluctuations is defined
by $D_{\sigma} (x, \tau) =
\langle T_{\tau} \left [ \sigma^{\dagger} (x, \tau ) \sigma ( 0,0 ) \right ]
\rangle$. We get $D_{\si}^{-1} (\bq, i \Om_n) = 1/J_K + \Pi_{fc} (\bq, i \Om_n)$,
where
\beq
\label{eq:pifc}
\Pi_{fc} (\bq, i \Om_n) =
\frac{1}{\be} \sum_{\bk, i \om_n} G_c (\bk, i \om_n)
G_f (\bk + \bq, i \om_n - i \Om_n)
\eeq
is the interband polarization bubble between the conduction and the spinon bands.
In the above $G_c^{-1} (\bk, i \om_n) = (i \om_n - \ep_{\bk})$ is the
propagator for the conduction electrons, while $G_f^{-1} (\bk, i \om_n) = (i \om_n - \ep^0_{\bk})$
is the propagator for the dispersive spinons. We write
$\Pi_{fc} (\bq, i \Om_n) = \Pi_{fc} (\bq, 0) + \Delta \Pi_{fc} (\bq, i \Om_n)$,
where $\Pi_{fc} (\bq, 0)$ is the static part of the fluctuations and
$\Delta \Pi_{fc} (\bq, i \Om_n)$ is the dynamic part. We first compute the static part
which can be written as
\beq
\label{eq:pifcq-1}
\Pi_{fc} (\bq, 0) = \sum_{\bk} \frac{n_F(\ep_{\bk}) - n_F(\ep^0_{\bk + \bq})}
{\ep_{\bk} - \ep^0_{\bk + \bq}},
\eeq
where $n_F(\ep)$ is the Fermi function.
We find that $\Pi_{fc}(\bq, 0)$ is independent of momentum
if the dispersions are linearized in Eq.~(\ref{eq:pifcq-1}). This implies that the momentum
dependence is due to $k \sim k_F$ in the $\bk$-integral of Eq.~(\ref{eq:pifcq-1}), for which
it is important to retain the quadratic dispersions of the bands. Furthermore, since the
main contribution is for $k \sim k_F$, the small momentum scale $q^{\ast}$ is unimportant
and can be set to zero to facilitate the calculation,
and we write $\ep_{\bk} = (k^2 - k_F^2)/(2m)$
and $\ep^0_{\bk} = (k^2 - k_F^2)/(2 m_0)$. Then, in terms of $\al$, the ratio of the two
bandwidths that we introduced earlier, we have $\al = m/m_0$.
Using $ \bk \leftrightarrow \bk + \bq$ inside the $\bk$-summation we get
\bea
\Pi_{fc}(\bq, 0)
&=&
\sum_{k \leq k_F} \left\{ \frac{1}{\ep_{\bk} - \ep^0_{\bk + \bq}}
- \frac{1}{\ep_{\bk + \bq} - \ep^0_{\bk}} \right\}
\nonumber \\
&=&
\int_0^{k_F} \frac{d k k^2}{4 \pi^2} \int_{-1}^{1} dz
\left\{ \frac{1}{B - \frac{qkz}{m}} - \frac{1}{C + \frac{qkz}{m_0}} \right\}
\nonumber \\
&=&
\int_0^{k_F} \frac{d k k}{4 \pi^2 q} \left[
m \ln \left| \frac{B + qk/m}{B - qk/m} \right| \right.
\nonumber \\
&-&
\left.
m_0 \ln \left| \frac{C + qk/m_0}{C - qk/m_0} \right| \right],
\nonumber
\eea
where $B = A (k_F^2 - k^2) - q^2/(2m)$, $C = A (k_F^2 - k^2) - q^2/(2m_0)$,
and $A = [1/(2m) - 1/(2m_0)]$.
After performing the momentum integration, we expand the resulting expression in powers of $(q/k_F)$,
and we use $\rho_0 = mk_F/(2 \pi^2)$, the density of states per spin of the conduction electrons at the Fermi level.
To leading order in $(q/k_F)$ we get
\bea
\label{eq:pifcq-2}
\Pi_{fc}(\bq, 0)
&=& \rho_0 \left[ \frac{\ln \al}{(1 - \al)} + \frac{1 - \al^2 + 2 \al \ln \al}{4 (1 - \al)^3}
\left(\frac{q}{k_F} \right)^2 \right]
\nonumber \\
&\approx&
\rho_0 \left[ - \ln \left( \frac{1}{\al} \right) + \frac{q^2}{4 k_F^2} \right].
\eea
Note that the $ \rho_0 \ln (\al)$ term in the above equation has been derived in appendix A2
using a slightly different method for the calculation of the mean field free energy
in Eq.~(\ref{eq:mf-free2}). This term, along with $1/J_K$, define the mass
$(1/J_K + \rho_0 \ln (\al))$ of the Kondo boson, which goes to zero at the QCP.

Next we calculate the dynamic part of the fluctuations which can be written as
\bea
\label{eq:Dpi-1}
\Delta \Pi_{fc}(\bq, i \Om_n)
&=& \Pi_{fc}(\bq, i \Om_n) - \Pi_{fc}(\bq, 0)
\nonumber \\
&=&
\frac{1}{\be} \sum_{\bk, \om_n} G_c (\bk, i \om_n)
\left[ G_f (\bk + \bq, i \om_n - i \Om_n)
\right. \nonumber \\
&-& \left. G_f (\bk + \bq, i \om_n) \right].
\eea
Unlike in the case of the static part, here the dominant contribution is
from the interband particle-hole excitations around the two Fermi surfaces,
for which the spectra can be linearized. We write $\ep_{\bk} = \ep$ for the
dispersion of the conduction electrons,
and $\ep^0_{\bk + \bq} = \al (\ep - v_F q^{\ast} + v_F q z)$ for the dispersion
of the spinons, where $z$ is the cosine of the angle between wavevectors $\bk$
and $\bq$. Approximating the $\bk$-summation by
\[
\sum_{\bk} \rightarrow \frac{\rho_0}{2} \int_{- \infty}^{\infty} d \ep \int_{-1}^{1} dz,
\]
at zero temperature we get
\begin{subequations}
\label{eq:Dpi-all}
\beq
\label{eq:Dpi-2}
\Delta \Pi_{fc}(\bq, i \Om_n) = \frac{\rho_0}{2(1- \al)} \left[
Y_1 + Y_2 + Y_3 + Y_4 \right],
\eeq
where
\beq
\label{eq:Y12}
Y_{1,2} = \left( 1 \mp \frac{E_1}{v_F q} \right) \ln (E_1 \mp v_F q)
\eeq
with $E_1 = v_F q^{\ast} - i \Om_n/\al$, and
\beq
\label{eq:Y34}
Y_{3,4} = - \left( 1 \mp \frac{E_2}{v_F q} \right) \ln (E_2 \mp v_F q)
\eeq
\end{subequations}
with $E_2 = v_F q^{\ast} - i \Om_n$. From the above expression of the
dynamic part given by Eqs.~(\ref{eq:Dpi-2})--(\ref{eq:Y34}), we next
extract the leading behaviour in different regimes of frequency and momentum.
For this we need to compare the momentum $q$ with $q^{\ast}$, and the
frequency $\Om_n$ (a continuous variable at
$T=0$) with the energy scales $E_x \equiv \al v_F q^{\ast}$ and $\al v_F q$.
Note that $v_F q^{\ast} \sim 10^3$ K is an energy  scale much larger than the
ultraviolet cut-off of the theory $\al D \sim 10$ K (the spinon bandwidth),
and therefore we need to consider only $\left| \Om_n \right| \ll v_F q^{\ast}$. We find
five distinct regimes which are as follows:
\newline
(i) $\left| \Om_n \right| < E_x$ and $q < q^{\ast}$, where
\begin{subequations}
\label{eq:Yis}
\beq
\label{eq:Yi1}
\sum_i^4 Y_i \approx -2 (1 - \al) \frac{i \Om_n}{E_x} \left[
1 + \frac{1}{3} \left(\frac{q}{q^{\ast}}\right)^2 + \frac{1}{2}
(1+ \al) \frac{i \Om_n}{E_x} \right].
\eeq
Note that in the above, we retained two sub-leading terms because
there are regimes where the sub-leading terms are larger than the
static $(q/k_F)^2$ term.
\newline
(ii) $\left| \Om_n \right| < E_x$ and $q > q^{\ast}$, where
\beq
\label{eq:Yi2}
\sum_i^4 Y_i \approx -2 (1 - \al) \frac{i \Om_n}{\al v_F q} \left[
i \frac{\pi}{2} {\rm sgn} (\Om_n) + \frac{q^{\ast}}{q} \right].
\eeq
(iii) $\left| \Om_n \right| > E_x$ and $q < q^{\ast}$, where
\beq
\label{eq:Yi3}
\sum_i^4 Y_i \approx 2 \left[ \ln \left( \frac{- i \Om_n}{E_x}
\right) + \frac{1}{6} \left(\frac{q}{q^{\ast}} \right)^2
- \frac{E_x}{i \Om_n} \right].
\eeq
(iv) $\left| \Om_n \right| > \al v_F q > E_x$ and $q > q^{\ast}$, where
\beq
\label{eq:Yi4}
\sum_i^4 Y_i \approx 2 \left[ \ln \left( \frac{- i \Om_n}{\al v_F q} \right)
+ 1 + i \frac{\pi}{2} {\rm sgn} (\Om_n) \right].
\eeq
(v) $\al v_F q > \left| \Om_n \right| > E_x$ and $q > q^{\ast}$, where
\beq
\label{Yi5}
\sum_i^4 Y_i \approx (1 - \al) \frac{i \Om_n}{\al v_F q} \left[
-i \pi {\rm sgn} (\Om_n) + (1 + \al) \frac{i \Om_n}{\al v_F q} \right].
\eeq
\end{subequations}
At the quantum critical point, the mass of the
Kondo boson goes to zero due to Eq.~(\ref{eq:JKc}).
The leading frequency and momentum dependences of $D_{\sigma} (\bq, i \Om_n)$ are
determined using Eqs.~(\ref{eq:pifcq-2}), (\ref{eq:Dpi-all}) and (\ref{eq:Yis}).
The details of the various asymptotic structures of $D_{\sigma} (\bq, i \Om_n)$
in different regimes of frequency and momentum are discussed in appendix \ref{appensub:Dsigma}.
Among the forms of $D_{\sigma} (\bq, i \Om_n)$ given in
Eqs.~(\ref{eq:Dsigma-1all})--(\ref{eq:Dsigma-3all}), only the following two asymptotic
structures are important for obtaining the leading contribution of the Kondo boson to
thermodynamic and transport properties.

First, for $\left| \Om_n \right| < [\al D/(2\pi)] (q^{\ast}/k_F)^3$ and
$q < q^{\ast}$,
we get
\beq
\label{eq:Dsigmaz2}
D^{-1}_{\si}(\bq, i \Om_n) \approx  \rho_0 \left[ \frac{1}{4} \left(
\frac{q}{k_F} \right)^2 - \frac{i \Om_n}{E_x} \right],
\eeq
which gives rise to an undamped propagating mode with dynamical exponent $z=2$ (the
dispersion of which is given by setting Eq.~(\ref{eq:Dsigmaz2}) to zero). The existence
of this mode is a direct consequence of the mismatch between the Fermi surfaces of
the conduction and the spinon bands. Due to this mismatch, a minimum momentum of
$q^{\ast}$ is necessary to excite an interband particle-hole pair. Consequently, for
momentum $q < q^{\ast}$, the spectrum of the Kondo boson lies outside the continuum
of the interband particle-hole excitations and thereby remains undamped.
Note that this massless mode corresponds to hybridization fluctuations about the QCP, and becomes
massive for $J_K < J_{Kc}$ (this is realized by adding a constant term $\delta$
to  Eq.~(\ref{eq:Dsigmaz2})).
Since $\Pi_{fc}$ at $q$=0 diverges logarithmically at $E_x$, the mode energy never exceeds $E_x$.
The mode dispersion, which is quadratic about $q$=0, is more complicated as $q$ approaches
$q^{\ast}$ due to logarithmic corrections to $\Pi_{fc}$,
and is described in greater detail in appendix \ref{appensub:Mode}.

Second, for most of the phase space,
the spectrum for the fluctuations
of $\si$ lies within the interband particle-hole continuum, and we get
\beq
\label{eq:Dsigmaz3}
D^{-1}_{\si}(\bq, i \Om_n) \approx  \rho_0 \left[ \frac{1}{4} \left(
\frac{q}{k_F} \right)^2 + \frac{\pi}{2} \frac{\left| \Om_n \right|}{\al v_F q} \right],
\eeq
i.e., an overdamped critical mode with dynamical exponent $z=3$.
Next we note that,
since we assume $q^{\ast} \ll k_F$, the overdamped $z=3$ critical mode occupies
most of the momentum space and therefore almost always it provides the leading contribution to
thermodynamic and transport properties. In this regime, the scaling of frequency is given by
$\Om_n \sim [(\al D)/(2 \pi)] (q/k_F)^3$, and since this regime ends for $q < q^{\ast}$, one obtains the
infrared energy scale
\beq
\label{eq:E-star}
E^{\ast} \approx c {\al D} \left( \frac{q^{\ast}}{k_F} \right)^3,
\eeq
where $c$ is $1/(2\pi)$.  The true value of $c$ is slightly smaller ($\sim$ 0.1) since there
are logarithmic corrections to $\Pi_{fc}$ as $q$ approaches $q^{\ast}$.  A more detailed
account is given in appendix \ref{appensub:Mode}.
We note that $E^{\ast}$, which can
be estimated to be $\sim$ 1 mK, appears as
an infrared crossover scale for any physical property that is affected by the excitations
of $\si$. On the other hand, the ultraviolet cutoff scale is provided by $\al D \sim 10$ K,
which is the bandwidth of the spinons, or equivalently
the single ion Kondo scale by Eq.~(\ref{eq:JhTk}).

\section{Thermodynamics of the Kondo boson}
\label{sec:thermodynamics}

In this section, we study the effect of the fluctuations of the Kondo boson $\si$ on the
thermodynamics of the system in the quantum critical regime. In particular, we compute
(a) the contribution to the free energy, (b) the temperature dependence of the static spin
susceptibility, and (c) the crossover lines in temperature which demarcate the quantum
critical regime.

\subsection{Free energy}
\label{subsec:free}

The contribution of the fluctuations of $\si$ to the free energy (per unit volume)
is given by
\beq
\label{eq:free-general}
F = \sum_{\bq} \int_{- \infty}^{\infty} \frac{d \Om}{2 \pi} \coth \left( \frac{\Om}{2 T} \right)
{\rm Im} \ln \left[ D_{\si}^{-1} (\bq, \Om + i \eta) \right],
\eeq
where $D_{\si}(\bq, \Om + i \eta)$ denotes the retarded propagator for the Kondo bosons. We find
that, for all temperatures $T < \al D$, the leading $T$ dependence of the free energy $F$ is given
by that part of phase space where the mode is overdamped (with dynamical exponent $z=3$) and for
which the
expression for the propagator is approximately given by Eq.~(\ref{eq:Dsigmaz3}). The details of this demonstration,
as well as the evaluation of the sub-leading contribution from the other regimes, is given in
appendix  \ref{appensub:free-energy}.
For $T > E^{\ast}$, the leading $T$ dependence of $F$ is given by
\bea
\label{eq:F2}
F &\approx&
- \frac{k_F^3 \al D}{2 \pi^3} \int_0^{\infty} d \Om
\coth \left( \frac{\Om}{2 T} \right)
\int_{q_c}^1 dq \, q^2
\nonumber \\
& \times &
\tan^{-1} \left( \frac{2 \pi \Om}{q^3} \right).
\eea
Here $q$ and $q_c$ are dimensionless momenta in units of $k_F$, and $\Om$ and
$T$ are dimensionless energies in units of $\al D$.
Since the $q$-integral is ultraviolet divergent, we use the Fermi momentum
as an upper cutoff.
The infrared cut-off, $q_c$, for the z=3 regime
is dependent on the particular temperature range considered, since
the leading $T$ dependence comes from frequencies $\Om \sim T $.
For $T > E^{\ast}$,
$q \gtrsim \Om^{1/3}$, for which we can approximate $\tan^{-1} (x) \approx x$ and replace
the cutoff $q_c$ by $\Om^{1/3}$.
Performing the integrals, we find
\beq
\label{eq:F3}
F(T) \approx - \left( \frac{k_F^3}{9} \right) \ln \left( \frac{\al D}{T} \right) \frac{T^2}{\al D},
\quad T > E^{\ast}.
\eeq
We note that this contribution adds to a similar $T^2 \ln (T)$ contribution from the transverse gauge
fluctuations (which are massless $z=3$ excitations).
They give rise to a $\ln(T)$ behavior for the specific heat coefficient.

For $T < E^{\ast}$, the leading contribution to
the free energy is again given by Eq.~(\ref{eq:F2}) with $q_c = q^{\ast}/k_F$ for the infrared cutoff of the
$q$-integral. This is because for $\Om \sim T < E^{\ast}$, the $z=3$ regime exists for $q > q^{\ast}$.
As a result, because $\Om^{1/3} < q^{\ast}/k_F$ in this
temperature regime, the lower cut-off remains at $q^{\ast}/k_F$. This gives,
\beq
\label{eq:F4}
F(T) \approx - \left( \frac{k_F^3}{3} \right) \ln \left( \frac{k_F}{q^{\ast}} \right) \frac{T^2}{\al D},
\quad T < E^{\ast}.
\eeq
This $T^2$-dependence cannot be distinguished from ordinary Fermi liquid corrections, and in this temperature
regime the free energy is dominated by the $T^2 \ln (T)$ contribution from the transverse gauge
fluctuations~\cite{reizer}.

\begin{figure}[tbp]
\begin{center}
\includegraphics[width=2.4in]{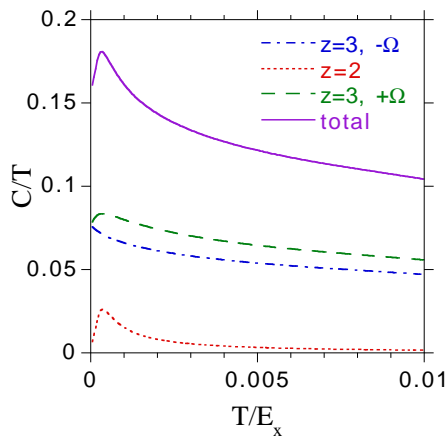}
\end{center}
\caption{(Color online) Numerical estimate of the contribution to the specific heat coefficient, C/T, coming
from the Kondo boson.  A value of $\alpha=0.001$ and $q^*/k_F=0.1$ was assumed, with the momentum integral
cut off at $k_F$ and the frequency integral at $0.1 E_x$.  Note the logarithmic behavior of
the z=3 contribution which is cut off for $T < E^*$ (with $E^* \sim 0.001 E_x$), and the sub-leading
nature of the z=2 contribution.}
\label{fig:spec}
\end{figure}

The collective mode gives a magnon-like contribution to the free energy ($F \sim T^{5/2}$), and is
sub-leading relative to the z=3 contribution (see appendix \ref{appensub:free-energy}).
We illustrate this by showing in Fig.~\ref{fig:spec} a numerical determination of the
contribution of the specific heat coefficient, C/T, coming from the Kondo boson,
using the fc polarization bubble of Eq.~(\ref{eq:Dpi-2}).  In this plot, one sees
the sub-leading contribution arising from the z=2 region, the logarithmic contribution from the z=3
region which saturates for $T < E^*$, and the small difference between the positive and
negative $\Omega$ contributions
from the z=3 region due to the chirality of the fc polarization bubble.

\subsection{Static spin susceptibility}
\label{subsec:spin-susc}

At the mean field level, where the critical fluctuations of $\si$ are ignored, the
temperature dependence of the static spin susceptibility $\chi_s (T)$ is entirely
analytic, namely a constant (Pauli susceptibility) plus a $T^2$ term, which is usual
for band fermions. Next, when we take the critical fluctuations into account, we expect the
correction to $\chi_s (T)$ to be non-singular (since the transition is non-magnetic and the
excitations of $\si$ are in the singlet channel), but non-analytic (due to the massless excitations).
In order to evaluate this temperature dependence, we first need to compute $D_{\si} (\bq, i \Om_n)$
in the presence of a magnetic field ($B$). For a finite $B$, the effect of the Zeeman term is to shift
the Fermi wave vectors $k_F$ and $k_F^0$ of the conduction and the spinon bands, respectively.
We get, $k_F^0 \rightarrow k_F^0 \pm (\mu_B g_f B)/(\al v_F)$, and
$k_F \rightarrow k_F \pm (\mu_B g_c B)/v_F$, where $g_f$ and $g_c$ are effective
 Lande $g$-factors of the
spinons and the conduction electrons, respectively, $\mu_B$ is the Bohr magneton, and $\pm$ refers to
the up and down spins, respectively. Since $\al \ll 1$, and in general $g_f > g_c$, we can ignore the
coupling of $B$ to the $c$-electrons and consider the effect of the Zeeman term as a renormalization of
the mismatch wave vector $q^{\ast}$, which is given by
\[
q^{\ast} \rightarrow q^{\ast} \pm \frac{\mu_B g_f B}{\al v_F}.
\]
Next, we note that, in the presence of a finite $q^{\ast}$, one expects $\Pi_{fc}(0,0)$ to have
corrections of the type $q^{\ast}/k_F$ and $(q^{\ast}/k_F)^2$ (which are not calculated in Eq.~(\ref{eq:pifcq-2}),
since the evaluation was performed in the limit $q^{\ast} \rightarrow 0$). This implies that,
in the presence of a magnetic field, we expect a correction to $\Pi_{fc}(0,0)$ which is
proportional to $[(\mu_B g_f B)/(\al D)]^2$ (since the excitation of $\si$ is in the singlet
channel, we do not expect a linear term in $B$). Adding such a term to $D_{\si} (\bq, i \Om_n)$,
and noting that the leading temperature dependence is due to the overdamped $z=3$ mode,
we can generalize Eq.~(\ref{eq:F2}) to obtain the $B$ dependence of the free energy as
\bea
\label{eq:FB}
F (B, T) &\approx&
- \frac{k_F^3 \al D}{2 \pi^3} \int_0^{\infty} d \Om
\coth \left( \frac{\Om}{2 T} \right)
\int_{q_c}^1 dq \, q^2
\nonumber \\
& \times &
\tan^{-1} \left( \frac{2 \pi \Om}{q^3 + h^2 q} \right).
\eea
Here energy and momenta are in dimensionless units (as in Eq.~(\ref{eq:F2})) and $h = (\mu_B g_f B)/(\al D)$ is the
dimensionless magnetic field. Writing the correction to the static spin susceptibility due to
the fluctuations of $\si$ as $\dl \chi_s (T) \equiv - [\partial^2 F/(\partial B)^2]_{B = 0}$,
we get for $T > E^{\ast}$
\beq
\label{eq:d-chi-s1}
\dl \chi_s (T) \approx
- (\mu_B g_f)^2 \left[\frac{2^{4/3} \Gamma (4/3) \zeta (4/3)}{\pi^{5/3} 3^{3/2}} k_F^3 \right]
\frac{T^{4/3}}{(\al D)^{7/3}},
\eeq
while for $T < E^{\ast}$, the lower cut-off is at $q^{\ast}$, making the mode
effectively massive, and we get
\beq
\label{eq:d-chi-s2}
\dl \chi_s (T) \approx
- (\mu_B g_f)^2 \left[ \frac{1}{3} \frac{k_F^5}{(q^{\ast})^2} \right]
\frac{T^2}{(\al D)^3}.
\eeq
As in the case of the free energy, the non-analyticity in the leading temperature dependence is
cutoff below $E^{\ast}$ due to the mismatch wave vector $q^{\ast}$.
As noted before, the gauge bosons give rise to a $T^2\ln(T)$ contribution to $\chi_s$.

\subsection{Crossover lines defining the quantum critical regime}
\label{subsec:cross}

The crossover lines in temperature that demarcate the quantum critical regime are symmetric
about the QCP $\dl = \dl_c = 0$, where $\dl = 1/(\rho_0 J_K) - 1/(\rho_0 J_{Kc})$ is the
dimensionless tuning parameter of the theory (for fixed $J_H$).
On the heavy Fermi liquid side of the QCP, such a line usually defines
the boundary of the finite temperature phase transition.
However, it has been argued in the
literature that
the gauge fluctuations convert the finite temperature mean field phase transition
line into a crossover line~\cite{senthil,nagaosa-lee}.
These lines are determined by the temperature dependent
mass $\dl m (T)$ of the excitations of $\si$.
 In a Ginzburg-Landau approach, these excitations are generated by the quartic $u_0 | \si |^4$
coupling in the action, where $u_0 \sim \rho_0/(\al D)^2$ from Eq.~(\ref{eq:mf-free2}).
 In the following, we compute
$\dl m (T)$ generated due to the propagating mode with $z=2$ given by Eq.~(\ref{eq:Dsigmaz2}),
as well as that generated by the overdamped mode with $z=3$ given by Eq.~(\ref{eq:Dsigmaz3}).
The contributions from the other regimes of $D_{\si} (\bq, i \Om_n)$ are always sub-leading.
The general expression for $\dl m (T)$ is given by
\beq
\label{eq:dm0}
\dl m (T) = u_0 \sum_{\bq} \int_{- \infty}^{\infty} \frac{ d \Om}{2 \pi} \coth \left(
\frac{\Om}{2 T} \right) {\rm Im} D_{\si} (\bq, \Om + i \eta).
\eeq
Denoting the contribution of the $z=2$ mode as $\dl m_1 (T)$, we get using Eq.~(\ref{eq:Dsigmaz2})
\[
\dl m_1 (T) = \left( \frac{u_0 E_x}{2 \pi^2 \rho_0} \right) \int_0^{q^{\ast}} dq \, q^2 \,
n_B \left( \frac{E_x q^2}{4 k_F^2} \right),
\]
where $n_B (x) = (e^{x/T} - 1)^{-1}$ is the Bose function.
For the leading $T$-dependence,
we write $n_b (x) \approx 1/x$, with an appropriate ultraviolet cutoff for the $q$-integral. For
$T < E^{\ast}$, this cutoff is $k_F (T/E_x)^{1/2}$, and for $T > E^{\ast}$, this cutoff remains at $q^{\ast}$.
We get
\beq
\label{eq:dm1}
\begin{split}
\dl m_1 (T)
& \approx
\left( \frac{4 u_0 k_F^3}{ \pi^2 \rho_0} \right) \frac{T^{3/2}}{E_x^{1/2}},
\quad T < E^{\ast},
\\
& \approx
\left( \frac{2 u_0 k_F^2 q^{\ast}}{ \pi^2 \rho_0} \right) T,
\quad T > E^{\ast}.
\end{split}
\eeq
Next, denoting the contribution of the $z=3$ mode as $\dl m_2 (T)$, we get
\bea
\dl m_2 (T)
&=& \left( \frac{4 u_0 \al D k_F^3}{\pi^2 \rho_0} \right)
\int_0^{\infty} d \Om \, \coth \left( \frac{\Om}{2 T} \right) \, \Om
\nonumber \\
&\times &
\int_{q_c}^{\infty} dq \frac{q^3}{q^6 + 4 \pi^2 \Om^2},
\nonumber
\eea
where $q$ and $q_c$ are dimensionless in units of $k_F$, and $\Om$ and $T$
are dimensionless in units of $\al D$.
For $T > E^{\ast}$, we can put
$q_c \sim \Om^{1/3}$ for the leading term, while for $T < E^{\ast}$ we have
$q_c = q^{\ast}$. This gives,
\beq
\label{eq:dm2}
\begin{split}
\dl m_2 (T)
& \approx
\left[ \frac{2 u_0 k_F^3}{3 \rho_0} \right] \left( \frac{k_F}{q^{\ast}} \right)^2
\frac{T^2}{\al D}, \quad T < E^{\ast},
\\
& \approx
\left[ \frac{2^{7/3} \Gamma (4/3) \zeta (4/3) u_0 k_F^3}{3^{3/2} \pi^{5/3} \rho_0} \right]
\frac{T^{4/3}}{(\al D)^{1/3}}, \quad T > E^{\ast}.
\end{split}
\eeq
Comparing Eqs.~(\ref{eq:dm1}) and (\ref{eq:dm2}) we find that, for $T < E^{\ast}$, the
leading $T$ dependence is given by the $z=2$ mode and $\dl m (T) \approx \dl m_1 (T)$,
while for $T > E^{\ast}$, the leading term is from the $z=3$ damped mode and $\dl m (T) \approx \dl m_2 (T)$.
Consequently, the crossover lines in temperature which define the quantum critical regime are given by
\beq
\label{eq:crossover-lines}
\begin{split}
T & \propto  \left| \dl - \dl_c \right|^{2/3}, \quad T < E^{\ast},
\\
& \propto  \left| \dl - \dl_c \right|^{3/4}, \quad T > E^{\ast}.
\end{split}
\eeq

\section{Quasiparticle lifetime and transport}
\label{sec:lifetime}

In this section, we first evaluate the quasiparticle lifetime ($\tau_{c}$) of the
conduction electrons due to scattering from the excitations of $\si$,
and then argue that this lifetime can be identified with the transport lifetime ($\tau_{tr}$)
for the evaluation of the temperature dependence of the resistivity.

\begin{figure}[tbp]
\begin{center}
\includegraphics[width=5.0cm]{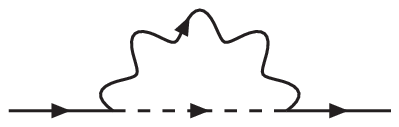}
\end{center}
\caption{Self-energy diagram for the conduction electrons due to scattering from
the critical excitations of $\sigma$ whose dynamical exponent is $z=3$. In three
dimensions, this has a marginal Fermi liquid form.}
\label{fig:selfenergy}
\end{figure}

For the process shown in Fig.~\ref{fig:selfenergy}, the general expression for the imaginary part
of the self-energy of the conduction electrons is given by
\begin{align}
\label{eq:self-general}
{\rm Im} \Sigma_c^R (\bk, \om)
& =
\sum_{\bq} \int_{- \infty}^{\infty} \frac{d \Om}{\pi}
\left[n_B(\Om) + n_F(\Om - \om)
\right]
 \notag \\
 & \times
{\rm Im} D_{\si}^R (\bq, \Om) \,
{\rm Im} G_f^R (\bk - \bq, \om - \Om),
\end{align}
where $n_F(x) = (e^{x/T} + 1)^{-1}$ is the Fermi function,
and $R$ denotes retarded functions. At zero temperature, this gives
\begin{align}
\label{eq:self-T0}
{\rm Im} \Sigma_c^R (\bk, \om)
& =
\sum_{\bq} \int_0^{\om} \frac{d \Om}{\pi} \, {\rm Im} D_{\si}^R (\bq, \Om)
\notag \\
& \times
{\rm Im} G_f^R (\bk - \bq, \om - \Om).
\end{align}
We evaluate the above expression for a conduction electron on the Fermi surface,
i.e., for $| \bk | = k_F$, and we find that the leading frequency dependence is
always due to the overdamped $z=3$ mode whose expression is given by
Eq.~(\ref{eq:Dsigmaz3}).
The $z=2$ mode does not contribute since it cannot
kinematically connect the $f$ and $c$ electrons.
We write
\[
\sum_{\bq} \rightarrow \frac{1}{4 \pi^2} \int_0^{\infty} dq \, q^2 \int_{-1}^{1} dz,
\]
where $z$ is the cosine of the angle between $\bk$ and $\bq$.
After linearizing the spectrum for the spinons, we have
\[
{\rm Im} G_f^R ({\bf k}_F - \bq, \om - \Om) = - \pi \dl (\om - \Om + E_x + \al v_F q z).
\]
Since, $\Om \sim \om$, and
$q > q^{\ast}$ for the overdamped mode,
the constraint from the $\dl$-function is always satisfied. After the angular integral, we
get
\[
{\rm Im} \Sigma_c^R (k_F, \om) = - \left( \frac{2 k_F^3}{\pi \rho_0} \right)
\int_0^{\om} d \Om \, \Om \int_{q^{\ast}}^{\infty} dq \frac{q^2}{q^6 + 4 \pi^2 \Om^2},
\]
where momenta and frequencies are dimensionless in units of $k_F$ and $\al D$,
respectively.
For $\Om \sim \om > E^{\ast}$, the leading contribution of the $q$-integral comes
from $q \sim (\Om)^{1/3}$ and therefore the infrared cutoff $q^{\ast}$ can be set to zero.
But for $\Om \sim \om < E^{\ast}$,
$q^{\ast} > \Om^{1/3}$, and the lower cut-off at  $q^{\ast}$ comes into play.
We finally get,
\beq
\label{eq:Sigma-omega-dep}
\begin{split}
{\rm Im} \Sigma_c^R (k_F, \om)
& \approx
- \left( \frac{k_F^3 }{6 \pi \rho_0 \al D} \right) \left| \om \right|,
\quad \left| \om \right| > E^{\ast},
\\
& \approx
- \left( \frac{k_F^3 }{6 \pi^2 \rho_0 \al D E^{\ast}} \right) \om^2,
\quad \left| \om \right| < E^{\ast}.
\end{split}
\eeq
Thus we find that above the infrared cutoff scale $E^{\ast}$, the Kondo breakdown scenario,
in which the conduction electrons interact with the critical hybridization fluctuations,
provides a microscopic mechanism to obtain a marginal Fermi liquid in three dimensions.

Next we evaluate the temperature dependence of the imaginary part of the self-energy
at $\om = 0$ and on the Fermi surface. Denoting this as ${\rm Im} \Sigma_c (T)$, we get
from Eq.~(\ref{eq:self-general})
\begin{align}
\label{eq:self-w0}
{\rm Im} \Sigma_c^R (T)
& =
\sum_{\bq} \int_{- \infty}^{\infty} \frac{d \Om}{\pi} \, \frac{1}{\sinh(\Om/T)} \, {\rm Im} D_{\si}^R (\bq, \Om)
\notag \\
& \times
{\rm Im} G_f^R ({\bf k}_F - \bq,  - \Om).
\end{align}
The evaluation of the above expression is very similar to the finite frequency case,
except for $T > E^{\ast}$, the thermal factors $n_B (x) + n_F (x) = 1/\sinh(x)$ gives an
additional logarithm which is cut off by $E^{\ast}$. We get,
\beq
\label{eq:Sigma-T-dep}
\begin{split}
{\rm Im} \Sigma_c^R (T)
& \approx
- \left( \frac{k_F^3 }{3 \pi \rho_0 \al D} \right) T \ln \left( \frac{2 T}{E^{\ast}} \right),
\quad T > E^{\ast},
\\
& \approx
- \left(  \frac{k_F^3 }{6 \rho_0 \al D E^{\ast}} \right) T^2,
\quad T < E^{\ast}.
\end{split}
\eeq

\begin{figure}[tbp]
\begin{center}
\includegraphics[width=2.4in]{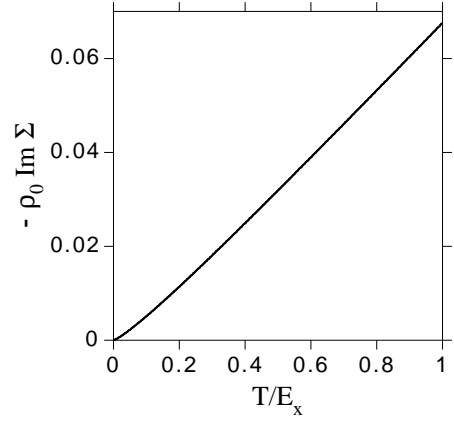}
\end{center}
\caption{The imaginary part of the conduction electron self-energy at $\omega=0$ versus $T$
from a numerical evaluation of Eq.~(\ref{eq:self-w0}) using Eq.~(\ref{eq:Dsigmaz3}).  Note the
approximate linear $T$ behavior, with a crossover to $T^2$ behavior at very low $T$.}
\label{fig:transp}
\end{figure}

In Fig.~\ref{fig:transp}, we show a plot of this quantity from a numerical evaluation of Eq.~(\ref{eq:self-w0}) using Eq.~(\ref{eq:Dsigmaz3}).  One can see the approximate linear $T$ behavior except at very low
temperatures, where one crosses over to a $T^2$ behavior.

\begin{figure}[tbp]
\begin{center}
\includegraphics[width=6.0cm]{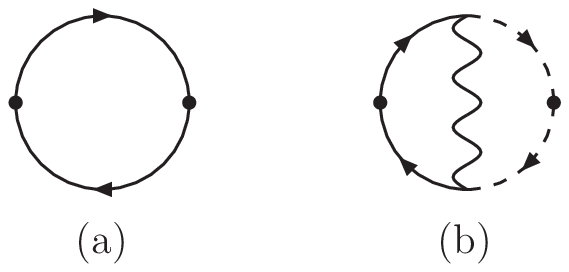}
\end{center}
\caption{Diagrams for the current-current correlator for the evaluation of the conductivity
in the Kubo formalism. Solid dots indicate current vertices. (a) Involves conduction electrons with
self-energy corrections [see Fig.~\ref{fig:selfenergy}]. This contribution identifies the transport
lifetime with the conduction electron lifetime.
(b) Vertex correction involving the exchange of one $\sigma$ boson.
It vanishes because the spinon current is zero due to the local
constraint [see Eq.~(\ref{eq:constraint})].  }
\label{fig:transport}
\end{figure}

Next we evaluate the temperature dependence of the resistivity $\dl \rho (T) \equiv \rho (T) - \rho (0)$. In order to proceed,
we first need to address whether the transport lifetime $\tau_{tr}$ can be identified with the quasiparticle lifetime
$\tau_c (\om, T) \propto [{\rm Im} \Sigma_c (\om, T)]^{-1}$, whose frequency and temperature dependences are given by
Eqs.~(\ref{eq:Sigma-omega-dep}) and (\ref{eq:Sigma-T-dep}). For this, it is useful to compare the Kondo-Heisenberg model
with a single band model. In the latter case, the two lifetimes have a different temperature dependence because the leading
contribution to the self-energy comes from forward scattering processes with momentum transfer $q \simeq 0$, but which
are not effective in relaxing the current. As such, when vertex corrections are taken into account, $\tau_{tr}^{-1}$
acquires an additional temperature dependence proportional to $q^2 \sim T^{2/z}$. However, this is not the case for the
Kondo-Heisenberg model which has two bands, one of light conduction electrons and the other of heavy spinons.
Due to the
constraint of half filling [Eq.~(\ref{eq:constraint})], the spinon current operator $\vec{J}_{f i} = 0$ at every site $i$.
Therefore, it is guaranteed by gauge invariance
that a vertex correction involving
the exchange of a single $\si$ boson [Fig.~\ref{fig:transport}b],
which involves an external spinon current operator, is identically zero~\cite{gauge}.
 The first non-zero vertex correction involves the
exchange of two $\si$ bosons, and we expect such a correction to be small by a factor of $\al$.
This can be understood as well in a Boltzmann approach~\cite{ziman}, where the transport
vertex correction
$1-\cos(\theta)$ gets replaced by $1-\alpha \cos(\theta)$, which is essentially unity since
$\alpha \ll 1$.

Consequently, in the present
theory, the transport lifetime is proportional to the quasiparticle lifetime. The physical picture that emerges from the above
discussion is that, when scattered from a $\si$ boson ($c \rightleftharpoons f + \sigma$), a conduction electron transmutes
into a spinon and relaxes its current in the bath of the spinons. More formally, the expression for the conductivity ($\si_c$) obtained
from the current-current correlator in the Kubo formalism is given by
\beq
\si_c = \left( \frac{v_F^2}{3} \right) \sum_{\bk} \int_{- \infty}^{\infty} \frac{d \om} {2 \pi}
\left[ \frac{\partial}{\partial \om} \tanh \left( \frac{\om}{2 T} \right) \right]
\left[ {\rm Im} G_c^R (\bk, \om) \right]^2.
\eeq
We write
\beq
\left[ G_c^R (\bk, \om) \right]^{-1} = \om - \ep_{k} - {\rm Re} \Sigma (\om) + \frac{i}{2 \tau_c (\om, T)},
\eeq
where for $(\om, T) > E^{\ast}$
\beq
\left[ \tau_c (\om, T) \right]^{-1} = \tau^{-1} +
\left( \frac{2 k_F^3 }{3 \pi \rho_0 \al D} \right) \left[
T \ln \left( \frac{2 T}{E^{\ast}} \right) + \frac{| \om |}{2} \right].
\eeq
Here $\tau$ is an elastic scattering lifetime of the conduction electrons due to impurities, and sets the
scale of the temperature independent part of $\si_c$. We linearize the dispersion of the conduction electrons
and replace the momentum sum by an energy integral, and we finally obtain
 $\si_c (T) = [\rho_0 v_F^2 \tau_{c}(0, T)]/3$. This implies for $E^{\ast} < T < \al D$,
\beq
\label{eq:resistivity}
\dl \rho (T) \propto T \ln \left( \frac{2 T}{E^{\ast}} \right).
\eeq
Therefore, the scenario of the breakdown of the Kondo effect captures one of the most enigmatic
features of heavy fermion systems close to quantum criticality, namely the quasi-linear temperature
 dependence of the resisitivity observed for most compounds over a large range of temperature.
For $T < E^{\ast}$, the usual Fermi liquid result is recovered and $\dl \rho (T) \propto T^2$.
It is interesting to note that, the recovery of the Fermi liquid $T^2$ behavior of resistivity
below a finite temperature scale in the quantum critical regime of YbRh$_2$Si$_2$
has recently been reported~\cite{flouquet}.
Finally, we note that for the same reason that equates the single particle and transport lifetimes,
the electrical and thermal transport lifetimes are the same.

\section{Conclusion}

To summarize, we studied the Kondo-Heisenberg model in three dimensions using a fermionic representation
for the localized spins.
The mean-field phase diagram in the $T-J_K$ plane, where $J_K$ is the
Kondo coupling, exhibits a quantum critical point that separates a uniform spin liquid phase from a heavy
Fermi liquid phase. In the uniform spin liquid phase, the Kondo hybridization between the conduction band and
the band of fermionic spinons that constitute the local moments vanishes, thereby indicating that in this phase,
the Kondo effect fails to occur. For a Kondo coupling larger than the critical value ($J_K > J_{K_c}$), a heavy
Fermi liquid ground state is established with finite hybridization between the bands. This implies that at the
quantum critical point ($J_K = J_{K_c}$), the lattice Kondo energy scale $T_K$ vanishes, indicating the breakdown
of the Kondo effect for couplings smaller than the finite value $J_{K_c}$.

In general, the size of the (hot) ``Fermi surface'' of the spinon band is different from that
of the conduction electrons, and we characterized their mismatch by a wavevector $q^{\ast} \equiv k_{F0} - k_F$,
where $k_{F0}$ is the spinon Fermi wavevector and $k_F$ is the conduction Fermi wavevector.
As a consequence of this mismatch,
we found that
two mean field solutions are possible in the Fermi liquid phase. First, one with a uniform hybridization,
which is stabilized when the two band masses have the same sign. This is the standard Kondo phase which appears in the
mean-field pseudofermion description of the Kondo lattice. Second, a novel Kondo phase with the hybridization modulated
in space with wavevector $q_0 \approx 1.2 q^{\ast}$, which appears when the two band masses have
opposite signs (i.e., one band is electron-like and the other hole-like). Conceptually, this phase is analogous to the
LOFF state of superconductivity, and is characterized by nodes in space where $T_K$ is zero. In this paper, we did not
examine the physical consequences of the modulated hybridization, which will be the topic for future work.
For the
uniform case, we showed that at the quantum critical point the single ion Kondo scale $T_K^0$ is approximately equal to
the Heisenberg coupling $J_H$. This demonstrates that the Kondo breakdown is a consequence of the competition between
the Kondo energy scale and the magnetic energy scale, even though there is no long range magnetic order in the present
formulation.

Then, we studied the effect of the critical hybridization fluctuations (excitations of the order
parameter $\si$) associated with the vanishing energy
scale $T_K$ on the thermodynamic and transport properties of the system. We found that, due to the mismatch $q^{\ast}$,
the critical fluctuations are affected by energy scales $E^{\ast} \sim [\al D/(2 \pi)](q^{\ast}/k_F)^3$ and
$E_x \sim \al v_F q^{\ast}$, where $\al \sim J_H/D$ is the ratio of the spinon bandwidth $J_H$ and the conduction
bandwidth $D$. The propagator for the critical modes has several asymptotic structures in different regimes of
frequency-momentum space, out of which the following two are important and readily understood.
(i)
For momentum $q < q^{\ast}$, the spectrum of the critical fluctuations lies outside the interband particle-hole
continuum and therefore their dynamics is undamped and is characterized by a dynamical exponent $z=2$. (ii)
For most of momentum space ($q > q^{\ast}$), the spectrum of the critical modes lies within the
particle-hole continuum, and therefore have overdamped dynamics with exponent $z=3$ (Landau damping). The leading
contribution to thermodynamics and transport is almost always governed by the latter asymptotic structure,
in contrast to most Ginzburg-Landau approaches, where only the critical modes within $1/\xi(T)$ of
the ordering vector ($q=0$) are important~\cite{GL}.
Above the
temperature scale $E^{\ast} \sim 1$ mK, this gives rise to anomalous metallic behaviour, such as a specific heat coefficient
that diverges logarithmically with temperature, and the inverse lifetime of the conduction electrons which has a
$T \ln T$ temperature dependence. The latter is a consequence of the conduction electrons scattering with the critical
bosons with the dynamical exponent $z=3$, which in three dimensions provide a microscopic mechanism to obtain marginal Fermi
liquid behaviour.
Since the spinons do not carry current, but are effective in relaxing the current carried
by the conduction electrons, the $T$-dependence of the inverse particle lifetime
also gives rise to a $T \ln T$ behaviour of the resistivity. From a scaling point of view, in this regime the frequency
$\Om$ of the critical fluctuations scale as $ \Om \sim q^3$, where $q$ is their momentum. For $T < E^{\ast}$, however, the
infrared cutoff $q^{\ast}$ prohibits the $z=3$ scaling, and the leading $T$ dependence of the specific heat coefficient
and the resistivity are Fermi liquid like.

The Kondo breakdown scenario is promising in that it can
explain one of the least understood features of the heavy fermions near quantum criticality, namely the quasi-linear
temperature dependence of the resistivity,
and the existence of multiple energy scales, over decades of temperature.

\begin{acknowledgments}

We thank the hospitality of the KITP where this work was initiated. We also
acknowledge J. Schmalian, P. Sharma,  A. Chubukov, P. Coleman, G. Kotliar, M. Civelli, and L. De Leo for illuminating
discussions. This work was supported by the U.S. Dept.~of Energy, Office of
Science, under Contract No.~DE-AC02-06CH11357, in part by the National
Science Foundation under Grant No.~PHY99-07949 and by the French ANR36ECCEZZZ.

\end{acknowledgments}

\appendix

\section{}
\label{appen:meanfield}

\subsection{Spatially modulated mean field solution}
\label{appensub:modulated}

Here we demonstrate that when the conduction electron and spinon masses have opposite signs, i.e.,
when one band is electron-like and the other hole-like, the mean field theory admits a solution
where the Kondo hybridization is modulated in space. This is a consequence of the mismatch between
the two Fermi surfaces, and conceptually is analogous to the LOFF state of superconductivity~\cite{fflo,rice}.
In the following, we choose the conduction band to be electron-like and the spinon band hole-like, and linearize
their dispersions. This gives $\ep_{\bk} = \ep$ for the dispersion of the
conduction band, where $\ep = v_F (k - k_F)$, and $\ep_{\bk}^0 = - \al (\ep - v_F q^{\ast})$ for the dispersion
of the spinon band. For this case, we evaluate the static interband polarization $\Pi_{fc}(\bq, 0)$,
whose general expression is given by Eq.~(\ref{eq:pifcq-1}). Approximating the momentum summation by
\[
\sum_{\bk} \rightarrow \frac{\rho_0}{2} \int_{- D}^{D} d \ep \int_{-1}^{1} dz,
\]
where the conduction bandwidth $D$ enters as an ultraviolet cutoff for the energy integral, we get,
\bea
\label{eq:pifc-eh}
\Pi_{fc}(\bq, 0) &=&
\frac{\rho_0}{1 + \al}
\left\{ \ln \left[ \frac{ \al v_F^2 \left| (q^{\ast})^2 - q^2 \right|}{(1 + \al)^2 D^2} \right]
-2
\right. \nonumber \\
&+& \left.
\frac{q^{\ast}}{q} \ln \left[ \frac{q^{\ast} + q}{\left| q^{\ast} - q \right| } \right]
 \right\}.
\eea
It is easy to see that the maximum of $-\Pi_{fc}(\bq, 0)$ is at a finite wavevector $q_0$ where
\beq
\label{eq:q0}
q_0 \approx 1.2 q^{\ast}.
\eeq
Therefore for $J_K > J_{K_c}$, where the critical value of the Kondo coupling is given by
\[
\frac{1}{J_{K_c}} + \Pi_{fc}(q_0, 0) = 0,
\]
the Kondo boson condenses in the Fermi liquid phase at a finite wavevector $q_0$
(i.e., $\langle \si_{q_0} \rangle \neq 0$). This implies
that the Kondo hybridization is modulated, with nodes in space where $T_K$ vanishes~\cite{foot2}.

\subsection{Calculation of the mean field free energy}
\label{appensub:free}

In this part, we give the technical details for the evaluation of the
mean field free energy at $T=0$. This can be written as
\beq
\frac{F_{MF}}{N} = \sum_{\bk, i = a, b} \ep_{\bk}^i \theta( - \ep_{\bk}^i )
+ \frac{\si_0^2}{J_K} + \frac{\phi_0^2}{J_H} + \frac{\lam_0}{2},
\eeq
where $\theta(x)$ is the Heaviside step function, and $\ep_{\bk}^{a,b}$ are given by
Eq.~(\ref{eq:epsilon-ab}). We replace $\sum_{\bk} \rightarrow \rho_0 \int d \ep$,
and from the solution of the equations $\ep_{\bk}^{a,b} = 0$, we get
\begin{widetext}
\bea
\sum_{\bk, i = a, b} \ep_{\bk}^i \theta( - \ep_{\bk}^i )
&=&
\frac{\rho_0}{2} \int_{-D}^{s - s_1} d \ep
\left[
(1+ \al) \left( \ep + \frac{\ep^2}{D} \right)
 - \left\{(1- \al)^2 \left( \ep + \frac{\ep^2}{D} \right)^2 + 4 \si_0^2
\right\}^{1/2}
\right]
\nonumber \\
&+&
 \frac{\rho_0}{2} \int_{-D}^{- s - s_1} d \ep
\left[
(1+ \al) \left( \ep + \frac{\ep^2}{D} \right)
 + \left\{(1- \al)^2 \left( \ep + \frac{\ep^2}{D} \right)^2 + 4 \si_0^2
\right\}^{1/2}
\right],
\nonumber
\eea
\end{widetext}
 where $s = \si_0/\al^{1/2}$ and
$s_1 = \si_0^2/(2D) + \mathcal{O}(1/D^2)$. We expand the expression under the
square root in powers of $(1/D)$ and keep terms up to $\mathcal{O} (1/D^2)$,
since higher orders contribute to $\mathcal{O} (\si_0^6)$ and beyond
which we neglect. Performing the $\ep$-integral, to $\mathcal{O} (\si_0^4)$
accuracy we get
\bea
\frac{F_{MF}}{N}
&=&
\frac{\rho_0 D^2}{2} \left[
\frac{\al^2}{2 \rho_0 J_H} - \frac{\al}{3} \right]
+ \rho_0 \si_0^2 \left[ \frac{1}{\rho_0 J_K}
\right. \nonumber \\
&-&
\left.
\frac{1}{1 - \al} \ln \left( \frac{1}{\al} \right) \right]
+ \frac{\rho_0 \si_0^4}{\al^2 D^2} \frac{(1- 4 \al + \al^2)}{(1+ \al)},
\nonumber
\eea
where a constant part has been ignored. Since $\al \ll 1$, in the terms
proportional to $\si_0^2$ and $\si_0^4$, we retain only the dominant
$\al$-dependence, and get Eq.~(\ref{eq:mf-free2}).

\section{}
\label{appen:fluctuations}

\subsection{Asymptotic structure of the Kondo boson}
\label{appensub:Dsigma}

In this appendix, we determine the leading frequency and momentum dependences of the
propogator for the Kondo boson $D_{\si}(\bq, i \Om_n)$ in the quantum critical regime
using Eqs.~(\ref{eq:pifcq-2}), (\ref{eq:Dpi-all}) and (\ref{eq:Yis}).
Its leading frequency dependence is given
by the first terms in Eqs.~(\ref{eq:Yi1})--(\ref{Yi5}), while the next term is determined
by comparing the static $(q/k_F)^2$ term in Eq.~(\ref{eq:pifcq-2}) with the sub-leading
terms of Eqs.~(\ref{eq:Yi1})--(\ref{Yi5}). The asymptotic structure of $D_{\si}(\bq, i \Om_n)$
in different regimes of frequency and momentum are as follows:
\newline
\emph{(1) $\Om_n \ll E^{\ast} \equiv [(\al D)/(2 \pi)] (q^{\ast}/k_F)^3$}.
In this frequency interval there are three sub-regimes depending on the magnitude of the momentum
$\bq$. We get
\newline
(a) $q \ll q_{\Om 1} \equiv [\Om_n/E_x] k_F$ (where $E_x \equiv \al v_F q^{\ast}$),
\begin{subequations}
\label{eq:Dsigma-1all}
\beq
D^{-1}_{\si}(\bq, i \Om_n) \approx - \rho_0 \left( \frac{i \Om_n}{E_x} \right)
\left[ 1 + \frac{1}{2} (1 + \al) \frac{i \Om_n}{E_x} \right],
\eeq
(b) $q_{\Om 1} \ll q \ll q^{\ast}$,
\beq
D^{-1}_{\si}(\bq, i \Om_n) \approx  \rho_0 \left[ \frac{1}{4} \left(
\frac{q}{k_F} \right)^2 - \frac{i \Om_n}{E_x} \right],
\eeq
(c) $q^{\ast} \ll q \ll k_F$,
\beq
D^{-1}_{\si}(\bq, i \Om_n) \approx  \rho_0 \left[ \frac{1}{4} \left(
\frac{q}{k_F} \right)^2 + \frac{\pi}{2} \frac{\left| \Om_n \right|}{\al v_F q} \right].
\eeq
\end{subequations}
\emph{(2) $E^{\ast} \ll \Om_n \ll E_x$}.
In this frequency interval there are four sub-regimes given by
\newline
(a) $q \ll q_{\Om 2} \equiv (\Om_n/E_x)^{1/2} q^{\ast}$,
\begin{subequations}
\label{eq:Dsigma-2all}
\beq
D^{-1}_{\si}(\bq, i \Om_n) \approx - \rho_0 \left( \frac{i \Om_n}{E_x} \right)
\left[ 1 + \frac{1}{2} (1 + \al) \frac{i \Om_n}{E_x} \right],
\eeq
(b) $q_{\Om 2} \ll q \ll q^{\ast}$,
\beq
D^{-1}_{\si}(\bq, i \Om_n) \approx - \rho_0 \left( \frac{i \Om_n}{E_x} \right)
\left[ 1 + \frac{1}{3} \left( \frac{q}{q^{\ast}} \right)^2 \right],
\eeq
(c) $q^{\ast} \ll q \ll q_{\Om 3} \equiv k_F [ (q^{\ast} \Om_n)/(\al k_F D)]^{1/4}$,
\beq
D^{-1}_{\si}(\bq, i \Om_n) \approx - \rho_0 \left( \frac{i \Om_n}{\al v_F q} \right)
\left[ i \frac{\pi}{2} {\rm sgn} (\Om_n) + \frac{q^{\ast}}{q} \right],
\eeq
(d) $q_{\Om 3} \ll q \ll k_F$,
\beq
D^{-1}_{\si}(\bq, i \Om_n) \approx  \rho_0 \left[ \frac{1}{4} \left(
\frac{q}{k_F} \right)^2 + \frac{\pi}{2} \frac{\left| \Om_n \right|}{\al v_F q} \right].
\eeq
\end{subequations}
\emph{(3) $E_x \ll \Om_n \ll \al D$}.
In this frequency range there are five sub-regimes given by
\newline
(a) $q \ll q_{\Om 4} \equiv q^{\ast} (E_x/\Om_n)^{1/2}$,
\begin{subequations}
\label{eq:Dsigma-3all}
\beq
D^{-1}_{\si}(\bq, i \Om_n) \approx  \frac{\rho_0}{(1- \al)}
\left[ \ln \left( \frac{- i \Om_n}{E_x} \right) - \frac{E_x}{i \Om_n} \right],
\eeq
(b) $q_{\Om 4} \ll q \ll q^{\ast}$,
\beq
D^{-1}_{\si}(\bq, i \Om_n) \approx  \frac{\rho_0}{(1- \al)}
\left[ \ln \left( \frac{- i \Om_n}{E_x} \right) + \frac{1}{6}
\left( \frac{q}{q^{\ast}} \right)^2 \right],
\eeq
(c) $q^{\ast} \ll q \ll q_{\Om 5} \equiv k_F [\Om_n/(\al D)]$,
\beq
D^{-1}_{\si}(\bq, i \Om_n) \approx  \frac{\rho_0}{(1- \al)}
\left[ \ln \left( \frac{- i \Om_n}{\al v_F q} \right) + 1 + i \frac{\pi}{2} {\rm sgn} (\Om_n)
\right],
\eeq
(d) $q_{\Om 5} \ll q \ll q_{\Om 6} \equiv k_F [\Om_n/(\al D)]^{1/2}$,
\beq
D^{-1}_{\si}(\bq, i \Om_n) \approx \rho_0 \left[
\frac{\pi}{2} \frac{\left| \Om_n \right|}{\al v_F q} - \frac{1}{2}(1 + \al) \frac{\Om_n^2}{(\al v_F q)^2}
\right],
\eeq
(e) $q_{\Om 6} \ll q \ll k_F$,
\beq
D^{-1}_{\si}(\bq, i \Om_n) \approx  \rho_0 \left[ \frac{1}{4} \left(
\frac{q}{k_F} \right)^2 + \frac{\pi}{2} \frac{\left| \Om_n \right|}{\al v_F q} \right].
\eeq
\end{subequations}

\subsection{Spectral response of the Kondo boson}
\label{appensub:Mode}

In the paper, several simplified expressions were used for the spectral response of the Kondo
boson.  Here, we give a more complete account.  The fc polarization bubble has some similarities
to the Lindhard function~\cite{lindhard}, but also differs from it in important respects.
 In particular, the particle-hole continuum of the Lindhard function exists for all
momenta, while this is not the case for the fc polarization as a result
of the mismatch between the conduction and spinon Fermi surfaces.
We have performed numerical calculations including the full quadratic dispersion of the fermions,
but they are very similar to results we
present here that are based on Eq.~(\ref{eq:Dpi-2}) plus the static curvature correction
(last term in Eq.~(\ref{eq:pifcq-2})).
The advantage of using Eq.~(\ref{eq:Dpi-2}) is that it is valid
for arbitrarily small $\alpha$.
All results here are for the retarded response function at T=0.
 We confine our discussion to the case where both
conduction and spinon bands have the same sign for the mass.

We begin with the fc bubble at $q=0$
\beq
Re\Pi_{fc}(0,\Omega) = \frac{\rho_0}{1-\alpha}\ln\frac{|\Omega-\alpha v_Fq^*|}{|\Omega-v_F q^*|}
\eeq
This expression contains two logarithmic singularities at the energies
$E_x \equiv \alpha v_F q^*$ and $v_F q^*$, where $q^* \equiv k_{F0} - k_F$ is the mismatch
vector between the conduction and spinon Fermi surfaces.
The imaginary part of $\Pi_{fc}$ is simply a step function with value $\pi\rho_0/(1-\alpha)$ between these
two energies.  Note that $\Pi_{fc}$ is not symmetric around zero energy (only the sum of it with
$\Pi_{cf}$ would be).  We have chosen $k_{F0} > k_F$.  For the reverse case, the singularities would
flip to the other side of the frequency axis.  The log singularity at $E_x$ plays an important role.
It guaranties that the Kondo boson propagator, $D \equiv J_K/(1 + J_K \Pi_{fc})$, always has a pole
between zero and $E_x$.  This pole is undamped since $Im\Pi_{fc}$ is zero below $E_x$.

\begin{figure}
\includegraphics[width=2.4in]{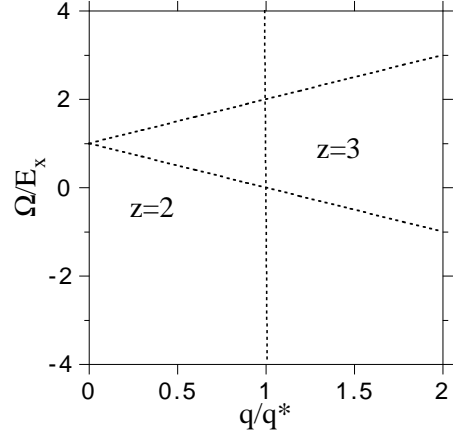}
\caption{Regimes of $\Pi_{fc}$.  The various dashed lines are the kinematic lines  corresponding
to the zeros of the arguments of the logs in Eq.~(\ref{eq:Dpi-2}).  The regime denoted $z=2$ has $Im\Pi_{fc} = 0$
whereas the regime $z=3$ has $Im\Pi_{fc} \propto \Omega$.}
\label{fig:domain}
\end{figure}

\begin{figure}
\includegraphics[width=2.4in]{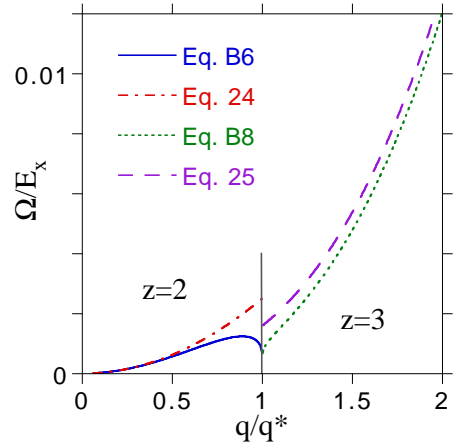}
\caption{(Color online) Dispersion of $ImD$ for $\delta=0$ and $q^*/k_F=0.1$.  The undamped ($z=2$) dispersion is to
the left of the kinematic boundary, $\Omega/E_x = 1-q/q^*$ (marked by the nearly vertical line), whereas
the damped ($z=3$) response is to the right.
The results based on Eq.~(\ref{eq:Dpi-2}) closely follow  the expressions of Eqs.~(\ref{eq:B6}) and (\ref{eq:B8}).
The simpler Eqs.~(\ref{eq:Dsigmaz2}) and (\ref{eq:Dsigmaz3})
are used in the analytic calculations, and are quite good except for $q$ near $q^*$.}
\label{fig:disp}
\end{figure}

The general structure of $\Pi_{fc}$ can be appreciated from Fig.~\ref{fig:domain}, where the various domains
for the imaginary part are shown.  Note that the imaginary part vanishes in the regime we label
as $z=2$.  For the positive frequency side, this is a triangle
in $(q, \Omega)$ space
bounded by $(0,E_x)$ and $(q^*,0)$.
For low frequencies appropriate for the dispersive peaks of $ImD$, it will be sufficient to expand
Eq.~(\ref{eq:Dpi-2}) for small $\Omega$.  When we do this, we find
\beq
Re D^{-1} = \delta - \frac{\rho_0\Omega}{2  \alpha v_F q}\ln\frac{|q+q^*|}{|q - q^*|}
+ \frac{\rho_0q^2}{4k_F^2}
\label{eq:B5}
\eeq
where $\delta$ is the deviation from the quantum critical point (QCP) and the last term is the static curvature
correction.
Below the kinematic boundary, $ImD^{-1}$ is zero,
so the zeros of Eq.~(\ref{eq:B5}) in this regime give the collective mode dispersion, which for $\delta=0$ is
\beq
\Omega_{coll}/E_x = 0.5(q^*/k_F)^2(q/q^*)^3/\ln\frac{|q+q^*|}{|q - q^*|}
\label{eq:B6}
\eeq
We compare this in Fig.~\ref{fig:disp}
to the expression where the log in Eq.~(\ref{eq:B5}) is expanded for small $q/q^{\ast}$,
the latter being Eq.~(\ref{eq:Dsigmaz2}).
Note that formally, Eq.~(\ref{eq:B6}) vanishes as $q$ goes to $q^*$, but this is of no concern,
since the mode intersects the kinematic boundary before this occurs, and thus
it terminates at a finite energy, corresponding to $c \sim 0.1$ in Eq.~(\ref{eq:E-star}).

Above the kinematic boundary, $ImD^{-1}$ is non-zero.
For $q > q^*$, it is
\beq
Im D^{-1} = \frac{-\rho_0\pi\Omega}{2\alpha v_F q}
\eeq
This leads to a pseudo-Lorentzian behavior for $ImD$.
The location of the maximum of $ImD$, denoted as $\Gamma$, can be found upon differentiation
with respect to $\Omega$, leading to
\begin{equation}
\Gamma/E_x = 0.5(q^*/k_F)^2(q/q^*)^3/\sqrt{\pi^2+\ln^2\frac{|q+q^*|}{|q - q^*|}}
\label{eq:B8}
\end{equation}
which is also plotted in Fig.~\ref{fig:disp}.  If instead, we ignore the $\Omega$
term in Eq.~(\ref{eq:B5}), we get Eq.~(\ref{eq:Dsigmaz3}) instead.  The latter is a true
Lorentzian, and its dispersion is plotted as well in Fig.~\ref{fig:disp}.
Although formally Eq.~(\ref{eq:B8}) vanishes as $q$ goes to $q^*$,
the actual results based on
Eq.~(\ref{eq:Dpi-2}) do not, and we again find $c \sim 0.1$ in Eq.~(\ref{eq:E-star}).

\begin{figure}
\includegraphics[width=3.4in]{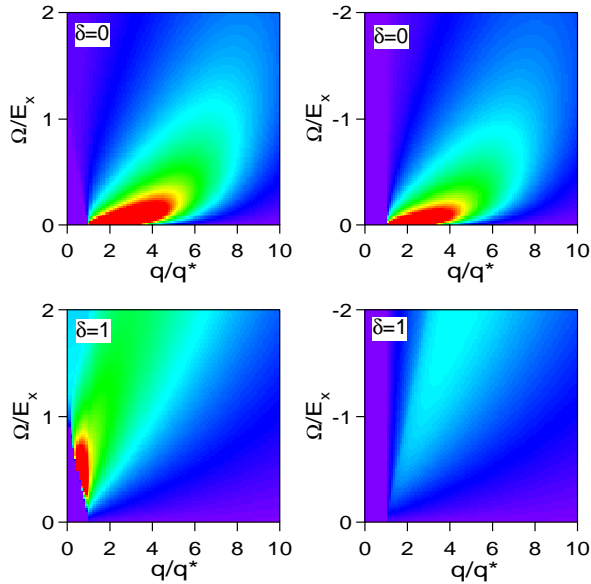}
\caption{(Color online) Plots of $ImD$ for positive (left) and negative (right) $\Omega$.
The quantum critical point ($\delta = 0$) is shown on the top, away from this ($\delta = 1$)
is shown on the bottom.  The $z=2$ dispersion is not visible on the scale of this plot.
Note the approximate (anti)symmetry of the damped ($z=3$) response at the QCP as
compared to away.  This damped dispersion at the QCP closely follows the analytic
expression of Eq.~(\ref{eq:B8}).
The intensity scale for the bottom plots are a factor of ten smaller than the top ones.}
\label{fig:boson}
\end{figure}

\begin{figure}
\includegraphics[width=2.4in]{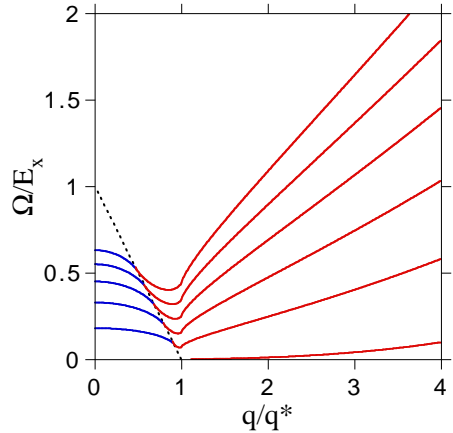}
\caption{(Color online) Dispersion of the $ImD$ maxima for $\delta$ ranging from zero (bottom curve) to
1 (top curve).  The undamped modes are to the left of the kinematic boundary (dashed line),
the damped modes to
the right.  Note the reversed magnon-like dispersion of the undamped modes
and the approximate linear $q$ behavior of the damped modes for non-zero $\delta$.}
\label{fig:omegaq}
\end{figure}

We finish this discussion by showing in Fig.~\ref{fig:boson} $ImD$ based on Eq.~(\ref{eq:Dpi-2}) for both positive and negative
$\Omega$ for two cases, the quantum critical point ($\delta = 0$) and somewhat away
($\delta = 1$).  The collective mode is not visible on the scale of this plot, but we note that
it is only present on the positive frequency side.  The damped response is approximately
(anti)symmetric in $\Omega$ for $\delta=0$ but becomes highly asymmetric for non-zero
$\delta$.  As $\delta$ increases, the most intense part of the damped response moves up the
kinematic boundary
 $\Omega/E_x = 1-q/q^*$ and approaches the log singularity at $q=0, \Omega=E_x$.
In Fig.~\ref{fig:omegaq}, the dispersion of the $ImD$ maxima is plotted for various
$\delta$.  Note the reversed magnon-like dispersion of the undamped modes
and the approximate linear $q$ behavior of the damped modes for non-zero $\delta$.

\subsection{Free energy}
\label{appensub:free-energy}

Here we compute the free energy due to the excitations of the Kondo boson, whose
expression is given by Eq.~(\ref{eq:free-general}), and take into account all the different
asymptotic structures of the propagator $D_{\si}(\bq, i\Om_n)$ which are given in
Eqs.~(\ref{eq:Dsigma-1all})--(\ref{eq:Dsigma-3all}). The goal of this exercise is to prove
that for all temperatures $T < \al D$, the leading contribution comes from that part of the
phase space where the boson is overdamped with dynamical exponent $z=3$,
and whose propagator is given by Eq.~(\ref{eq:Dsigmaz3}).

\emph{(1) $T < E^{\ast}$}. Since for the leading $T$ dependence we expect $\Om \sim T$, in this
temperature regime $D_{\si}(\bq, i\Om_n)$ has three asymptotic forms which are given in
Eq.~(\ref{eq:Dsigma-1all}). Accordingly, we split the $q$-integral into three parts,
namely $q < q_{\Om 1}$, $q_{\Om 1} < q < q^{\ast}$, and $q^{\ast} < q < k_F$, and denote
their contributions as $F_{1a}$, $F_{1b}$ and $F_{1c}$ respectively. Keeping only the
leading terms for each sub-regime, we get
\begin{widetext}
\begin{subequations}
\bea
F_{1a} &=&
\frac{1}{4 \pi^3} \int_{- \infty}^{\infty} d \Om \coth \left(
\frac{\Om}{2 T} \right) \int_0^{q_{\Om 1}} dq \, q^2
\, {\rm Im} \ln \left[ - \frac{\Om}{E_x} - i \eta \right]
= - \left( \frac{\pi^2}{90} k_F^3 \right) \frac{T^4}{E_x^3},
\\
F_{1b} &=&
\frac{1}{4 \pi^3} \int_{- \infty}^{\infty} d \Om \coth \left(
\frac{\Om}{2 T} \right) \int_{q_{\Om 1}}^{q_{\ast}} dq \, q^2
\, {\rm Im} \ln \left[ \frac{q^2}{4 k_F^2} - \frac{\Om}{E_x} - i \eta \right]
= - \left( \frac{\zeta (5/2)}{\pi^{3/2}} k_F^3 \right) \frac{T^{5/2}}{E_x^{3/2}},
\\
\label{eq:F1c}
F_{1c} &=&
\frac{1}{4 \pi^3} \int_{- \infty}^{\infty} d \Om \coth \left(
\frac{\Om}{2 T} \right) \int_{q_{\ast}}^{k_F} dq \, q^2
\, {\rm Im} \ln \left[ \frac{q^2}{4 k_F^2} - i \frac{\pi}{2} \frac{\Om}{\al v_F q} \right]
= - \left( \frac{k_F^3}{3} \right) \ln \left( \frac{k_F}{q^{\ast}} \right) \frac{T^2}{\al D}.
\eea
\end{subequations}
\end{widetext}
We note that, since $T < E^{\ast}$, the leading temperature dependence is due to the
$z=3$ mode whose contribution is given by Eq.~(\ref{eq:F1c}), and thus $F \approx F_{1c}$.
\newline
\emph{(2) $E^{\ast} < T < E_x$}.
In this temperature regime, $D_{\si}(\bq, i\Om_n)$ has four asymptotic forms which are given in
Eq.~(\ref{eq:Dsigma-2all}). Now we split the $q$-integral into four parts,
namely $q < q_{\Om 2}$, $q_{\Om 2} < q < q^{\ast}$, $q^{\ast} < q < q_{\Om 3}$ and $q_{\Om 3} < q < k_F$, and denote
their contributions as $F_{2a}$, $F_{2b}$, $F_{2c}$ and $F_{2d}$, respectively. Once again, keeping only
the leading terms for each sub-regime, we get
\begin{widetext}
\begin{subequations}
\bea
F_{2a} &=&
\frac{1}{4 \pi^3} \int_{- \infty}^{\infty} d \Om \coth \left(
\frac{\Om}{2 T} \right) \int_0^{q_{\Om 2}} dq \, q^2
\, {\rm Im} \ln \left[ - \frac{\Om}{E_x} - i \eta \right]
= - \left( \frac{\zeta (5/2)}{4 \pi^{1/2}} k_F^3 \right) \left( \frac{E^{\ast}}{\al D E_x^{3/2}} \right) T^{5/2},
\\
F_{2b} &=&
\frac{1}{4 \pi^3} \int_{- \infty}^{\infty} d \Om \coth \left(
\frac{\Om}{2 T} \right) \int_{q_{\Om 2}}^{q^{\ast}} dq \, q^2
\, {\rm Im} \ln \left[ - \frac{\Om}{E_x} - i \eta \right]
= - \frac{(q^{\ast})^3}{6 \pi^2} T \ln \left( \frac{T}{E^{\ast}} \right),
\\
F_{2c} &=&
\frac{1}{4 \pi^3} \int_{- \infty}^{\infty} d \Om \coth \left(
\frac{\Om}{2 T} \right) \int_{q_{\ast}}^{q_{\Om 3}} dq \, q^2
\, {\rm Im} \ln \left[- \left( \frac{q^{\ast}}{q} \right) \frac{\Om}{\al v_F q} - i \frac{\pi}{2}
\frac{\Om}{\al v_F q} \right]
\nonumber \\
&=& - \left( \frac{\Gamma (7/4) \zeta (7/4)}{6 \pi^2} k_F^3 \right) \left( \frac{q^{\ast}}{k_F} \right)^{3/4}
\frac{T^{7/4}}{(\al D )^{3/4}},
\\
\label{eq:F2d}
F_{2d} &=&
\frac{1}{4 \pi^3} \int_{- \infty}^{\infty} d \Om \coth \left(
\frac{\Om}{2 T} \right) \int_{q_{\Om 3}}^{k_F} dq \, q^2
\, {\rm Im} \ln \left[ \frac{q^2}{4 k_F^2} - i \frac{\pi}{2} \frac{\Om}{\al v_F q} \right]
= - \left( \frac{k_F^3}{9} \right) \ln \left( \frac{\al D}{T} \right) \frac{T^2}{\al D}.
\eea
\end{subequations}
\end{widetext}
After comparing the various contributions above, once again we find that the leading temperature
dependence is due to the $z=3$ mode, whose contribution is given by Eq.~(\ref{eq:F2d}), and we have
$F \approx F_{2d}$.
\newline
\emph{(3) $E_x < T < \al D$}.
In this temperature regime $D_{\si}(\bq, i\Om_n)$ has five asymptotic forms which are given in
Eq.~(\ref{eq:Dsigma-3all}). Now we split the $q$-integral into five parts,
namely $q < q_{\Om 4}$, $q_{\Om 4} < q < q^{\ast}$, $q^{\ast} < q < q_{\Om 5}$,
$q_{\Om 5} < q < q_{\Om 6}$ and $q_{\Om 6} < q < k_F$, and denote
their contributions as $F_{3a}$, $F_{3b}$, $F_{3c}$, $F_{3d}$ and $F_{3e}$ respectively. Once again, keeping only
the leading terms for each sub-regime, we get
\begin{widetext}
\begin{subequations}
\bea
F_{3a} &=&
\frac{1}{4 \pi^3} \int_{- \infty}^{\infty} d \Om \coth \left(
\frac{\Om}{2 T} \right) \int_{0}^{q_{\Om 4}} dq \, q^2
\, {\rm Im} \ln \left[ \ln \left( - \frac{\Om}{E_x} - i \eta \right) - \frac{E_x}{\Om} \right]
= - \left( \frac{(q^{\ast})^3}{18 \pi^2} \right) T,
\\
F_{3b} &=&
\frac{1}{4 \pi^3} \int_{- \infty}^{\infty} d \Om \coth \left(
\frac{\Om}{2 T} \right) \int_{q_{\Om 4}}^{q^{\ast}} dq \, q^2
\, {\rm Im} \ln \left[ \ln \left( - \frac{\Om}{E_x} - i \eta \right) + \frac{1}{6}
\left( \frac{q}{q^{\ast}} \right)^2 \right]
= - \left( \frac{(q^{\ast})^3}{12 \pi^2} \right) T \ln \left( \frac{T}{E_x} \right),
\\
F_{3c} &=&
\frac{1}{4 \pi^3} \int_{- \infty}^{\infty} d \Om \coth \left(
\frac{\Om}{2 T} \right) \int_{q^{\ast}}^{q_{\Om 5}} dq \, q^2
\, {\rm Im} \ln \left[ \ln \left( \frac{|\Om|}{\al v_F q} \right)
+ 1 - i \frac{\pi}{2} {\rm sgn} (\Om) \right]
= - \left( \frac{\pi^2}{90} k_F^3 \right) \frac{T^4}{(\al D)^3},
\\
F_{3d} &=&
\frac{1}{4 \pi^3} \int_{- \infty}^{\infty} d \Om \coth \left(
\frac{\Om}{2 T} \right) \int_{q_{\Om 5}}^{q_{\Om 6}} dq \, q^2
\, {\rm Im} \ln \left[ \frac{\Om^2}{2 (\al v_F q)^2} - i \frac{\pi}{2} \frac{\Om}{\al v_F q} \right]
= - \left( \frac{\zeta (5/2)}{8 \pi^{3/2}} k_F^3 \right) \frac{T^{5/2}}{(\al D )^{3/2}},
\\
\label{eq:F3e}
F_{3e} &=&
\frac{1}{4 \pi^3} \int_{- \infty}^{\infty} d \Om \coth \left(
\frac{\Om}{2 T} \right) \int_{q_{\Om 6}}^{k_F} dq \, q^2
\, {\rm Im} \ln \left[ \frac{q^2}{4 k_F^2} - i \frac{\pi}{2} \frac{\Om}{\al v_F q} \right]
= - \left( \frac{k_F^3}{9} \right) \ln \left( \frac{\al D}{T} \right) \frac{T^2}{\al D}.
\eea
\end{subequations}
\end{widetext}
As before, we find that the leading temperature dependence is given by the $z=3$ mode,
whose contribution is given by Eq.~(\ref{eq:F3e}), and we have $F \approx F_{3e}$.


\begin{thebibliography}{99}
\bibitem{foot1} Heavy fermion metals have complex Fermi surfaces where both conduction
and f surfaces are large.  As a consequence, large values of $q^{\ast}$ are unlikely.
Very small values of $q^{\ast}$ are possible due to degeneracies.  Within our simple model, a
value of $q^{\ast}/k_F \sim 0.1$ is a reasonable estimate.
\bibitem{gauge} As we showed in Section \ref{subsec:kondofluc},
gaussian fluctuations about the mean field solution generate a dispersion for the $\sigma$
field. As a consequence of gauge invariance, this also implies a coupling between the $\sigma$ and the gauge fields
which leads to a back-flow current. From the point of view of an effective field theory, where the $\sigma$
fields are dressed, the constraint $J_{f,i}  = 0$ is generalized to
$J_{f,i} + J_{\sigma, i} = 0$, and leads to an additional contribution to the conductivity
via the Ioffe-Larkin composition rules~\cite{ioffe}.
This contribution, though, is of order $\alpha$, and therefore does
not change our findings, since $\alpha \ll 1$~\cite{anderson}.
\bibitem{GL}  Note that the $z=2$ modes make sub-leading contributions to all quantities but
the crossover lines.  They also make no contribution to the conduction electron self-energy
and transport due to kinematic constraints.
\bibitem{foot2} For spherical Fermi surfaces, as many ordering
wave vectors as allowed by lattice symmetry will condense,
each with a modulus of $q_0$.
\end{thebibliography}
\end{document}